\documentclass[12pt,preprint]{aastex}

\usepackage{graphics,graphicx,epsf,natbib}
\newcommand{\lbol}{\mbox{$L_{\rm X}$}}
\newcommand{\tx}{\mbox{$T_{\rm X}$}}
\newcommand{\mvir}{\mbox{$M_{\rm vir}$}}

\newcommand{\nh}{N_{\rm H}}

\received{}

\shorttitle{Entropy Profiles} 
\shortauthors{Donahue, Horner, Cavagnolo and Voit}

\begin{document}

\title{Entropy Profiles in the Cores of Cooling Flow Clusters of Galaxies}

\author{ Megan Donahue\altaffilmark{1}, 
Donald J.  Horner\altaffilmark{2}, Kenneth W. Cavagnolo\altaffilmark{1}, and G. Mark
  Voit\altaffilmark{1}}

\altaffiltext{1}{Department of Physics and Astronomy, Michigan State
  University, BPS Building, East Lansing, MI 48824;
  donahue@pa.msu.edu, cavagnolo@pa.msu.edu, voit@pa.msu.edu}

\altaffiltext{2}{Code 660, Goddard Space Flight Center, Greenbelt, MD, horner@milkyway.gsfc.nasa.gov}

\begin{abstract}
The X-ray properties of a relaxed cluster of galaxies are determined primarily by its gravitational potential well and the entropy distribution of its intracluster gas.  That entropy distribution reflects both the accretion history of the cluster and the feedback processes which limit the condensation of intracluster gas.  Here we present {\em Chandra} observations of the core entropy profiles of nine classic ``cooling-flow" clusters  that appear relaxed and contain intracluster gas with a cooling time less than a Hubble time.  We show that those entropy profiles are remarkably similar, despite the fact that the clusters range over a factor of three in temperature.  They typically have an entropy level of $\approx 130 \, {\rm keV \, cm^2}$ at 100~kpc that declines to a plateau $\sim 10 \, {\rm keV \, cm^2}$ at $\lesssim 10$~kpc.  Between these radii, the entropy profiles are $\propto r^\alpha$ with $\alpha \approx 1.0 - 1.3$.  The non-zero central entropy levels in these clusters correspond to a cooling time $\sim 10^8 \, {\rm yr}$, suggesting that episodic heating on this timescale maintains the central entropy profile in a quasi-steady state.
\end{abstract}

\keywords{catalogs -- galaxies: clusters: general -- X-rays:
  galaxies: clusters -- cosmology: observations -- methods: data
  analysis}

\section{Introduction}\label{sec:intro}

The global properties of a cluster of galaxies, such as its bolometric X-ray luminosity \lbol\  and its mean temperature \tx\ , are determined primarily by the mass \mvir\ within a suitably chosen virial radius.  A cluster's temperature depends on mass because mass determines the depth of the cluster's potential well.  Its X-ray luminosity depends on mass because mass determines both the total number of baryons in the cluster and the potential well confining those baryons.  
However, several secondary factors combine to produce a dispersion in both \lbol\  and \tx\  at a fixed \mvir, and understanding the nature of that dispersion is crucial to doing precision cosmology with clusters.  One of those factors is merger shocks, which can temporarily raise both the luminosity and best-fitting temperature of a cluster \citep[e.g.,][]{2002ApJ...577..579R}.  A second is the shape of the potential well, because clusters whose potentials are more centrally concentrated tend to have higher central temperatures \citep[e.g.,][]{voit02}.  A third factor is the amount of intracluster gas with a cooling time less than the age of the universe.  The presence of such gas leads to both a large peak in the central surface brightness of a cluster and a central temperature gradient that rises with radius.  Consequently, clusters having larger amounts of gas with a short cooling time tend to have higher \lbol\  and lower \tx\  at a given value of \mvir \citep{1998MNRAS.297L..57A, 1994MNRAS.267..779F, markevitch98}.

Such clusters have often been called ``cooling-flow clusters," because the central gas was thought to condense and flow toward the center of the cluster as it radiated away its thermal energy \citep[see][for a recent review]{2004cgpc.symp..144D}.  Observations from {\em Chandra} and {\em XMM-Newton} now show that the central gas is not simply cooling to low temperatures and condensing in the manner originally envisioned \citep[e.g.,][]{Peterson2001, 2003ApJ...590..207P}.  Some form of feedback apparently prevents the central gas from condensing and forming stars, thereby truncating the high end of the galaxy luminosity function.  The nature of that feedback is currently an active topic of both observational and theoretical research, focusing largely on the role of outflows from active galactic nuclei in cluster cores.  

This paper analyzes archival {\em Chandra} data on nine cooling-flow clusters seeking clues to what keeps that gas from condensing and why clusters of a given mass have different amounts of gas with a short central cooling time.  The tactic we take in our analysis is to focus on the entropy profiles of these clusters.  We concentrate on entropy because it is a more fundamental property of the intracluster medium itself than either temperature or density alone. For example, the temperature of a cluster's gas primarily reflects the cluster's potential well depth; heating or cooling of the gas merely causes it to expand or contract in the potential well with only a modest change in temperature. The density of that gas depends on how much gravity can compress it in the cluster's potential well, and it is the specific entropy of the gas that determines its density at a given pressure.  Thus, the observable X-ray properties of a relaxed cluster of galaxies depend almost entirely on two physical attributes: (1) the shape and depth of the cluster's dark-matter halo, and (2) the entropy distribution of the intracluster gas.  
\citep[e.g.,][]{voit02}. 

Intracluster entropy is also intimately related to the cooling and feedback processes that govern galaxy evolution and that may also play a role in limiting condensation in cluster cores.  Theories and simulations of cluster formation which ignore these processes fail to reproduce the observable properties of present-day clusters.  If gravity alone were responsible for shaping the appearances of clusters and groups, then we would expect their properties to be nearly self-similar, with a luminosity-temperature relation like $L \propto T^2$.  Furthermore, we would expect groups and clusters to have similar surface-brightness profiles, when scaled to the virial radius of the system.  However, observations indicate that $L \propto T^{2.6-2.9}$ \citep{es91a,markevitch98} and that the surface-brightness profiles of groups are shallower than those of clusters \citep{hms99, pcn99}.

Many papers in the literature have attributed these deviations from self-similarity to an early episode of preheating by supernovae or active galaxies.  Heat input that establishes a uniform minimum entropy level in the intergalactic medium breaks self-similarity because the extra entropy makes that gas harder to compress as it accretes into dark-matter haloes \citep[e.g.,][]{kaiser91,eh91}.  Thus, gas in the smaller potential wells of groups would be less centrally concentrated than gas in the larger potential wells of clusters, thereby reducing the luminosities of groups relative to clusters and flattening their surface-brightness profiles. Such a model predicts that the central entropy profile would flatten into a core reflecting the minimum entropy level of the IGM.  However, several analyses have suggested that the energy input required to explain the observed relations through global preheating is implausibly extreme.

\cite{vb01} have argued that the entropy scale responsible for similarity breaking is not a global property of the intergalactic medium but rather a requirement set by radiative cooling---the observed entropy at the core radii of groups and clusters turns out to be similar to the entropy level at which intracluster gas would cool within a Hubble time.  Gas below this cooling threshold is therefore subject to cooling, condensation, and whatever feedback follows from that condensation, which may include both supernova and AGN activity.  In this scenario, cooling and feedback conspire to deplete the amount of gas below the cooling threshold: some of the low-entropy gas condenses and feedback subsequently raises the entropy of the remaining gas until both cooling and feedback shut down.  \cite{voit02} have built upon this principle to create cluster models that account for many observable X-ray properties, including the \lbol\ - \tx\ relation, the \mvir\ - \tx\ relation, the surface-brightness profiles of clusters, and their temperature gradients.  However, those models do not explain why clusters differ in the amount of gas that still remains below the cooling threshold.

Cluster entropy profiles have been presented by previous workers, with some of the earliest work coming from ROSAT and ASCA 
\citep{pcn99,2000MNRAS.315..689L}. They pointed out that if gravitational collapse was the only physical process affecting the intracluster gas, all entropy profiles should be self-similar. The temperature and density profiles available from ROSAT and ASCA
data were very limited, however. Results from XMM \citep{2005A&A...433..101P} show that outside their core radii, clusters
are similar, but the XMM data for inside the core radius of a cluster are limited by the spatial resolution of the telescope.

In order to better understand the sub-threshold gas, the processes that keep some of it from condensing, and the effects of those processes on the global X-ray properties of clusters, we are developing a {\em Chandra} library of intracluster core entropy distributions.  We selected the first batch of clusters for the library using criteria designed to ensure high-quality radial entropy profiles.  The number of X-ray events had to be sufficient for extraction of high-quality spectra in multiple annular regions, and the clusters had to be relaxed enough to have an unambiguous central emission peak.   We therefore compiled a master list of clusters available in the {\em Chandra} public archive as of December 1, 2003, when this project began, and cross-correlated it with the NASA/IPAC Extragalactic Database (NED) to obtain redshifts and positional data for the clusters.  We then added any observations noted as cluster observations (those with sequence numbers beginning with `8') that did not contain NED clusters.  After compiling this master list, we created {\em Chandra} images of all of the clusters and examined them to eliminate any clusters with insufficient counts for our study or with obvious substructure, such double peaks.  We also eliminated observations badly affected by background flares.  

This selection process yielded a sample consisting entirely of nearby ``cooling flow" clusters, most of which are considered classic examples of that category.  Table~\ref{tab:sample} lists that sample of nine clusters, including the cluster name, {\em Chandra} observation identification numbers, coordinates, maximum extraction radius, the redshift of the cluster used in spectral fits, its X-ray luminosity, and its mean X-ray temperature.  These last two quantities are taken from ASCA observations analyzed by D. Horner in his PhD thesis.\footnote{D. Horner's catalog and Ph.D. thesis are on-line at http://lheawww.gsfc.nasa.gov/\~horner/thesis.html.}

That is why this paper describing the first installment in our {\em Chandra} entropy library, which will ultimately contain clusters with a greater variety of core properties, focuses on the entropy profiles of classic cooling-flow clusters.  In \S\ref{sec:analysis} we present our data analysis methods and techniques, and \S\ref{sec:results} presents our results.   We find that the entropy profiles for the nine cooling-flow clusters in our sample are remarkably similar, suggesting that the process preventing condensation somehow maintains the entropy profiles in a quasi-steady state. In \S\ref{sec:discussion} we discuss those results in the context of some recent theoretical models, and  \S\ref{sec:summary} summarizes the paper.  Throughout the analysis, we assume a cosmology with $H_{0}$ = 70 km s$^{-1}$ Mpc$^{-1}$, $\Omega_{m} = 0.3$, and $\Omega_{\Lambda} = 0.7$.  The derived data products for the
Chandra Cluster Entropy Library such as the X-ray spectra, associated response files, and surface brightness profiles will be 
available through the NASA High Energy Space Archive (HEASARC), in 
the Chandra section of W3Browse\footnote{http://heasarc.gsfc.nasa.gov/W3Browse/chandra}.

\section{Data Analysis}\label{sec:analysis}

The goal of our analysis was to derive entropy profiles as a function of radius for the nine clusters in our sample, with entropy quantified in terms of the adiabatic constant $K = kT n_e^{-2/3}$.  Thus, we needed to determine the electron density $n_e$ and the gas temperature $T$ as functions of radius.  To do that, we first processed and cleaned the data as described in \S\ref{sec:data}.  Then, we divided each cluster into concentric annuli and extracted a spectrum in each annulus as described in \S\ref{sec:extraction}.  We describe our analysis of the projected spectra from gas within these annuli in \S\ref{sec:projected}.  In order to obtain gas temperature as a function of physical rather than projected radius, we then performed a deprojection analysis of the projected spectra as described in \S\ref{sec:deprojected}.  The deprojected cluster temperatures are not much different from the projected gas temperature for
these clusters, presumably because the radial emission profiles are so steep. 
The resulting temperature gradients are rather coarse, with the number of radial bins ranging from four to thirteen.  
Because we desired a more finely-grained representation of the density gradients, we determined them by deprojecting the exposure-corrected source counts in the 0.5-2.0~keV band as described in \S\ref{sec:nelec_sbprofiles}. 
In all the analysis, we use only the observations by the Advanced CCD Imaging Spectrometer (ACIS)
backside illuminated S3 chip, which generally constrained the analysis to the inner $\sim 100$~kpc for these nearby clusters.

\subsection{Data Processing} \label{sec:data}

We reprocessed the Level 1 events files from the Chandra archive with {\tt CIAO 3.2.2} and {\tt CALDB
  3.1.0} to obtain processed (Level 2) events files.   We applied updated gain
maps and standard grade filtering  to these events.  
Because emission from low-redshift clusters usually fills the whole S3 detector, 
we constructed matching blank-sky background event files for each 
observation, using Maxim Markevitch's blank
sky background database.
We cleaned the cluster data of flare contamination, following Maxim Markevitch's
cookbook, in order to match his background maps. 
This process involved inspecting a light curve for events between 2.5--7.0 keV and
removing time intervals with significant background flares, i.e., peaks
with count rates $\gtrsim$ 3 $\sigma$ and/or a factor of $\gtrsim$ 1.2
off the mean background level of the observations, using Maxim
Markevitch's light curve cleaning routine {\tt
  lc\_clean.sl}.\footnote{Available at
  http://hea-www.harvard.edu/$\sim$maxim/axaf/acisbg/}  We then created a final
events file for the purposes of studying the extended cluster emission 
by excluding any bright point sources near the cluster. The 
point sources were identified by a visual inspection of the image.

\subsection{Spectral Extraction} \label{sec:extraction}

In each cluster, we selected a region for spectral extraction by centroiding the
cluster emission and placing it at the center of a circular aperture 
with the largest radius that would fit entirely within the chip.
Because many clusters are centered at the aim point of the S3 chip
rather than the center of the chip, that maximum radius can be as small
as $\approx 2\arcmin$ (as opposed to $\approx 4\arcmin$ for the center
of the chip).  For many of the clusters, this is $\lesssim 0.1
r_{500}$, where $r_{500}$ (the radius at which the average cluster
density is 500 times the critical density) is roughly the virial
radius of the cluster.  We report the maximum radius in units of 
arcminutes and kiloparsecs in Table~\ref{tab:sample}.

We then divided this aperture into concentric annuli, determining the annular
boundaries from background-subtracted, cumulative count 
profiles (0.3-8.0 keV) by fixing the number of counts in each annulus to
be at least 10,000 - 20,000 counts.  The minimum number of total counts 
was chosen to allow us to test 
whether and where two-temperature spectral fits might be preferred over
single temperature fits, and to allow for a  simple two-parameter fit to the temperature
gradient.

We then extracted spectra for each annulus, along with corresponding 
blank-sky background spectra, using the {\em acisspec} script.  Two detector response 
files must be created for each individual spectrum, an Redistribution Matrix
File (RMF) and an Ancillary Response File (ARF), which take into account
the variability and the spatial dependencies across the detector.
As of {\tt CIAO 3.2} and {\tt CALDB 3.0}, the {\tt CIAO} tool {\em mkacisrmf} must be
run after spectra are extracted for each cluster using the {\em acisspec} script. The purpose
of {\em mkacisrmf} is to generate an RMF using a CCD spectral response with 
two components. One component does not include charge-transfer inefficiency (CTI) 
effects; the second incorporates the spatial variation in the chip response caused by  CTI. 
The weight map created by {\em acisspec} is utilized by {\em mkacisrmf} 
to make a count-weighted RMF. Finally the weighted ARF is created with {\em mkwarf}, using the same weight map
and the new weighted RMF, in order to match the energy grid of the weighted
RMF. (The last step utilizing {\em mkwarf} is only necessary because we fit our
data using XSPEC.) For these older data on extended sources from the  ACIS S3 chip, 
this new procedure results in temperatures and
metallicities only marginally different from that obtained with {\tt CALDB 2.29},
{\tt CIAO 3.0}. Therefore, one can be confident that, for the 
intents of this paper, the Chandra calibration has stabilized for archival data and our results are
robust to the details of how the RMF or ARF were computed.

However, in our experience since the beginning of this project,  
the evolution of the Chandra calibration software and the associated calibration data files
has occasionally led to significant changes in the best-fitting spectral parameters for these same observations since 
their original publication.  We discuss those changes in 
\S~\ref{sec:literature}. As of late 2005, such changes appear to be a thing of the past, as the calibration of the ACIS S3 detector seems to be converging to consistent results.

For the spectral fitting, we grouped the counts into bins with a minimum of 25 counts per bin, and 
restricted the fit to 0.7-7.0 keV.

\subsection{Projected X-ray Spectra} \label{sec:projected}

Once we had extracted the spectra, we fit them with plasma 
emission models using the software package 
 {\tt XSPEC 11.3.1} for several classes of input assumptions, including one-temperature (1-T) and 
 two-temperature (2-T) MEKAL models.  We fit each
annular spectrum between $0.7-7.0$ keV  individually, so that all free parameters are independent in
each annulus. Earlier versions of the {\em Chandra} calibration seemed to suggest that two-temperature
models were required.  However, after the {\em Chandra} calibration improved, 
we discovered that two-temperature models were no longer required to fit the vast majority  
of the cluster spectra in our sample. 
We also explored the effect of fixing the hydrogen column density ($\nh$) of X-ray absorbing gas
to the Galactic value in the direction of the cluster, finding that discrepancies between the best-fit hydrogen column density and the assumed Galactic value resulted in a systematic shift in the
derived cluster temperature.  Again, with earlier versions of the calibration data and software, 
this source of uncertainty affects the hotter clusters more than 
the moderate temperature clusters. As the calibration improved over the years we worked on 
this project, the impact of hydrogen column density anomalies on the best-fit temperature diminished.  As of the last two calibration versions, the best-fit $\nh$ is consistent with 
Galactic $\nh$ from \citet{DickeyLockman1990}. 
Therefore, our spectral results are reported for fits for which $\nh$ was constrained to an assumed Galactic value.

Each one-temperature fit takes four parameters: a temperature $T$, a heavy-element abundance fraction $Z$ relative to the solar values, an absorbing column density $N_{\rm H}$, and a spectral normalization defined to be
\begin{equation}
N = \frac{10^{-14}}{4 \pi d_A^2 (1+z)^2} \int n_e n_H dV
\end{equation}
where $n_e$ and $n_{\rm H}$ are the number densities of electrons and hydrogen nuclei, respectively,  in units of cm$^{-3}$, and $d_A$ is the angular-size distance in cm. 
Table~\ref{tab:1Tprfixed} gives the best fits for each annulus when $N_{\rm H}$ is fixed to the Galactic value toward the cluster taken from  \citet{DickeyLockman1990}.
The columns in Table~\ref{tab:1Tprfixed} 
are as follows:
\begin{enumerate}
  
\item Cluster name
  
\item Outer radius of annulus in arcminutes
  
\item $N_{\rm H}$ in units of $10^{20}$ cm$^{-3}$  
	
\item Best-fitting temperature in keV with 90\% confidence limits 
  
\item Heavy-element abundance $Z$ in solar units along with 90\% confidence limits 
          [The solar value of Fe/H is taken to be $4.68 \times 10^{-5}$ \citep{angr}.]
  
\item Spectral normalization $N$ of the MEKAL model fit
  
 \item Statistical uncertainty of the spectral normalization $N$ with 90\% confidence limits
  
\item Reduced $\chi^{2}$ value of the fit
  
\item The number of degrees of freedom in the fit
  
\end{enumerate}

In some cases, the $\chi^{2}$ values of the 1-T fits to the projected data are poor.  
A poor fit could mean either that multiple temperature components are present at a given radius or that the projected
spectrum in that annulus comes from gas of different temperatures because of a radial 
temperature gradient. To resolve this issue,  we deprojected the annular
spectra for each cluster.

\subsection{Spectral Deprojection} \label{sec:deprojected}

In a cluster with a centrally peaked surface-brightness profile, the spectrum from 
an annulus of a given projected radius is dominated by gas at radii similar 
to the projected radius but includes cluster emission along the entire line of sight 
through the cluster. In order to recover the properties of the gas as
a function of physical radius, one must correct for this projected emission.
To deproject spectra, one starts at the outermost projected annulus, fits an emission
model and then iteratively removes the contribution of the outer layers from inner annuli. 
This procedure requires, at minimum, some sort of assumption about the symmetry of
the cluster.  For this work, we assumed that the cluster emission is spherically symmetric.

To deproject the cluster data, we fit the spectra using the built-in deprojection model 
{\tt projct} in the software package XSPEC, tested and validated by  \citet{2005MNRAS.356..237J}.  
The fit simultaneously includes the projected 
spectra from all annuli in the cluster, and the initial guesses to the fit parameters were 
taken from the fits to the projected spectra.  
Table~\ref{tab:table1Tdeproj} lists our best-fit parameters when $N_{\rm H}$
is fixed at the Galactic value.  The columns in this table are the same as for 
Tables~\ref{tab:1Tprfixed}, except that the normalization in column (6) is
that for a spherical shell, not for the individual spectrum at that annulus.
The $\chi^2$ values quoted represent a single fit over all the spectra for the cluster instead
of those for the individual annuli.

The deprojected fits occasionally showed signs of instability in which the temperature
and metallicity would oscillate, with large uncertainties, from shell to shell. 
Since an accurate measurement of metallicity requires more counts than an accurate 
measurement of temperature, we tied the metallicities of neighboring annuli
together. The metallicity was constrained to be equal across groups of two or three annuli to make 
a common metallicity estimate at lower spatial resolution than temperature or normalization.
Even so, the best fit projected temperatures occasionally  exhibited instability from 
shell to shell. However, the excursions were smaller than the statistical uncertainty in the 
temperatures.

We also noticed that deprojections of spectra with background corrections based
on the deep field resulted in the outermost bin having somewhat higher 
normalizations (see also \citet{2005MNRAS.356..237J}). The outermost bin in most cases is
contaminated by emission originating even farther from the center of the cluster. To 
estimate the effect of cluster emission outside the outermost annuli, we deprojected
the spectral datasets for each cluster where 
the spectrum from the outermost annulus was background-corrected based on local background 
instead of the deep fields. This choice reduced the best-fit normalization in that bin but did not
change the best-fit temperature in that bin. Our results in this paper are therefore not sensitive to the
treatment of this outermost bin. We report the deprojection results for deep-background
subtracted data only in Table~\ref{tab:table1Tdeproj}, except for the case of Abell 1795
which had a higher local background than average. (The temperature for the 
second most outer annulus had no upper limit for A1795 data for which all
annuli had a deep-background correction.) The outer temperature was statistically
the same for either method.

Using this deprojection algorithm,  we obtained reasonable $\chi^2$ values for fits 
with a single-temperature plasma within each spherical shell. We tested the effect of including
a second temperature component, and occasionally a second component in the central sphere 
improved the $\chi^2$ somewhat, but not significantly. 
The three clusters for which there may be evidence in our data for
a second component are 2A0335+096, Abell 252, and Abell 2052. We 
report the results of a deprojection analysis where we allowed a second
thermal component in the innermost bin in Table~\ref{tab:table2Tdeproj}. 
The metallicity of this inner sphere
was constrained to be the same in both components.

We did not require second components in as many cases as previous workers, particularly
those who included soft energy bins in their fit ($E<0.7-0.8$ keV). This discrepancy is 
due to the immaturity of the Chandra calibration early in the mission. We found over the history of our own analysis that the need for a second temperature component decreased as  
the CALDB version number increased.  We will discuss the sensitivity of  
our derived entropy profiles to the presence or absence of a soft component in the central cores
in \S\ref{sec:entropy}.  

\subsection{Electron Densities from Deprojected 
Surface Brightness Profiles} \label{sec:nelec_sbprofiles}

Electron density profiles can be derived with much higher resolution than the temperature and abundance profiles given in Tables~\ref{tab:1Tprfixed}-\ref{tab:table1Tdeproj} 
because the count rate in a limited bandpass is much more sensitive to electron density
that it is to temperature.  We therefore derived high-resolution density profiles 
for each cluster by dividing our apertures into annuli of 2.5 and 5.0 arcseconds,
and using interpolated normalization-to-count-rate ratios to solve for the electron density 
corresponding to a given projected count rate.   

We created differential surface-brightness profiles from 0.5-2.0 keV to minimize
their dependence on temperature. We corrected the counts in each annular bin for vignetting 
and small variations in the net exposure time for each bin by extracting a exposure 
profile from a normalized exposure map created
assuming a mono-energetic photon spectrum of 1 keV. These corrections were typically less
than 5\% per bin. Comparisons with maps made for 0.5 and 1.5 keV photons showed that, 
the systematic uncertainty induced by assuming mono-energetic photons is negligible.
We found that the ratio between the 
spectral normalization quantity $N$ and the count rate in this bandpass (0.5-2.0 keV)  was relatively 
insensitive to temperature, changing by about 10-15\% over the full range of temperature in 
a given cluster. The conversion was actually more sensitive to metallicity than temperature
in this bandpass. Therefore, the derived electron density profiles have only limited
sensitivity to the details of how we interpolate cluster temperatures or even how the
cluster temperatures themselves were determined.

A deprojected emission profile (count rate per unit volume) was then computed for each 
cluster using a standard technique \citep{KCC1983}.
The deprojected emission profile was converted into electron densities using
the appropriate ratio of  a MekaL spectral normalization to spectral count rate in the same bandpass
as the surface brightness profile. This conversion takes into account both temperature and
abundance variations that somewhat affect the emissivity of the gas in the 0.5-2.0 keV 
energy range.

\section{Results}\label{sec:results}

Here we describe the results of our data analysis.  We find that the clusters in our sample have rising temperature gradients and declining abundance gradients, in agreement with previous work.  We briefly discuss those results in \S\ref{sec:abund}.  Then, in \S\ref{sec:entropy}, we present high-resolution entropy profiles for our clusters derived by interpolating the observed temperature gradients onto our density-profile bins.   These entropy profiles turn out to show a striking regularity that is best fit with a power law in radius plus a constant entropy pedestal.  We conclude the section with a cluster-by-cluster comparison of our results with previous observations of these clusters.

\subsection{Abundance and Temperature Profiles} \label{sec:abund}

All the clusters in our sample have temperature profiles that rise with radius.  In order to quantify those temperature gradients, we fit the deprojected temperature data for the seven clusters that had at least four annuli with a minimum of 20,000 counts with the power-law relation
\begin{equation}
  T(r) = T_{100} \left( \frac {r_{\rm mid}} {100~\rm{kpc}} \right)^{\alpha_T}
\end{equation}
where $T_{100}$ is the temperature in keV at 100 $h_{70}^{-1}$ kpc and $r_{\rm mid}$ is the midpoint between the inner and outer radii of the annulus.  We report the results of these fits in
Table~\ref{tab:fit-temp}. This simple form gave adequate fits for all clusters except Abell~2052. 
We find that the typical power-law index for these temperature profiles is $\alpha_T \approx 1/3$ inside $4\arcmin$, in agreement with the results of \citet{vf04}.  

The abundance measurements in these clusters, which are dominated by the iron lines in the spectrum, generally decline with radius.  The innermost regions tend to have Fe/H equal to 50-100\% the solar value, decreasing to 30\% solar at $\gtrsim 100$~kpc.  This finding generally agrees with previously observed trends in cooling-flow clusters \citep[e.g.,][]{dm01}.  In \S\ref{sec:literature} we discuss our abundance results for individual clusters in more detail.  

\subsection{Entropy Profiles} \label{sec:entropy}

Our primary new results concern the entropy gradients of cooling-flow clusters.  Using the spectral fitting and deprojection results of \S\ref{sec:analysis}, we derived a radial entropy profile for each cluster under the assumption that the temperature and density distributions were spherically symmetric.  Strict spherical symmetry is obviously an idealization of reality that applies better to some clusters than to others, but we wanted to derive entropy profiles for all the clusters in a uniform way.  To obtain the entropy profiles, we linearly interpolated temperature profiles onto the fine radial grid used for the surface-brightness deprojection in \S\ref{sec:nelec_sbprofiles}.  That gave us $n_e(r)$ and $T(r)$, from which we constructed $K(r) = kT(r)n_e^{-2/3}(r)$.  In order to test the effects of different binning schemes, we used bin widths of both $2.5\arcsec$ and $5\arcsec$ per annulus.  
Since deprojecting the spectra often did not give significantly different results from simpler analysis of projected
spectra, and since projected temperature profiles were better behaved, we used the projected temperatures 
for the profile fits reported in Tables~\ref{tab:fit-5arcsec} and
\ref{tab:fit-2.5arcsec}. To show how much of an effect this choice had on our results, we also include the 
results for entropy profiles obtained by using the power-law fits to  
$T(r)$ from deprojected spectra (Table~\ref{tab:fit-2.5arcsec-tdeproj}). 

Because interpolation of temperature within the innermost temperature bin is not well constrained, we treated the temperature gradient in a few ways, in order to probe the sensitivity of our results to how we handle the temperature profile. 
Method 1 modeled the temperature gradient in the innermost bin with a linear extrapolation from the adjacent bin.  Method 2 used a constant temperature within the innermost bin.  This assumption may have the effect of inducing a core in the entropy profile. The differences between the two methods provide an estimate of the systematic uncertainties in the modelling of the temperature structure in the innermost bin. Method 3, where we used the best-fit power law to the deprojected temperatures, provides yet another estimate which also minimizes the central entropy core since the temperature at $r=0$ is forced to be $T=0$ in such fits.  This third model should be thought of as a providing a lower limit on the central entropy, as there is no spectroscopic evidence for significant amounts of gas with such low temperatures in the centers of clusters.

We fit the entropy profiles derived using Methods 1-3 with two different formulae.  The first formula assumes that the entropy profile is a power law in radius plus a constant entropy pedestal $K_0$: 
\begin{equation}
 K(r) = K_0 + K_{100} \left( \frac {r} {100~\rm{kpc}} \right)^{\alpha} \; \; .
\end{equation} 
The second formula sets $K_0 = 0$, making the entropy profile a pure power law:
\begin{equation}
 K(r) = K_{100} \left( \frac {r} {100~\rm{kpc}} \right)^{\alpha} \; \; .
\end{equation}
Table~\ref{tab:fit-5arcsec} reports the best fits obtained with these two formulae using surface-brightness bins $5\arcsec$ in width.  Table~\ref{tab:fit-2.5arcsec} reports the best fits using surface-brightness bins $2.5\arcsec$ in width.  Table~\ref{tab:fit-2.5arcsec-tdeproj}
reports the best fits using surface-brightness bins $2.5\arcsec$ in width and the best fit to the deprojected temperature profile.
Our tables also report the range (in Mpc) of the fitted region. 

We used 1000 Monte-Carlo bootstrap reproductions of the original surface brightness data to quantify the statistical uncertainties of the deprojection process. The outermost deprojected bins in our fit are noisy and were excluded from the fit.  For one cluster, Abell 2052, the central surface brightness deprojection is strongly influenced by the presence of cavities and edges. The overall fits for this cluster consequently have large $\chi^2$ values because of these non-axisymmetric structures, which make the azimuthally averaged surface-brightness profile and therefore the deprojected density profile non-monotonic at $r \sim 10$~kpc.

\begin{figure*}
\includegraphics[width=7in,angle=0, trim = 0.7in 1.5in 0.65in 1.6in, clip ]{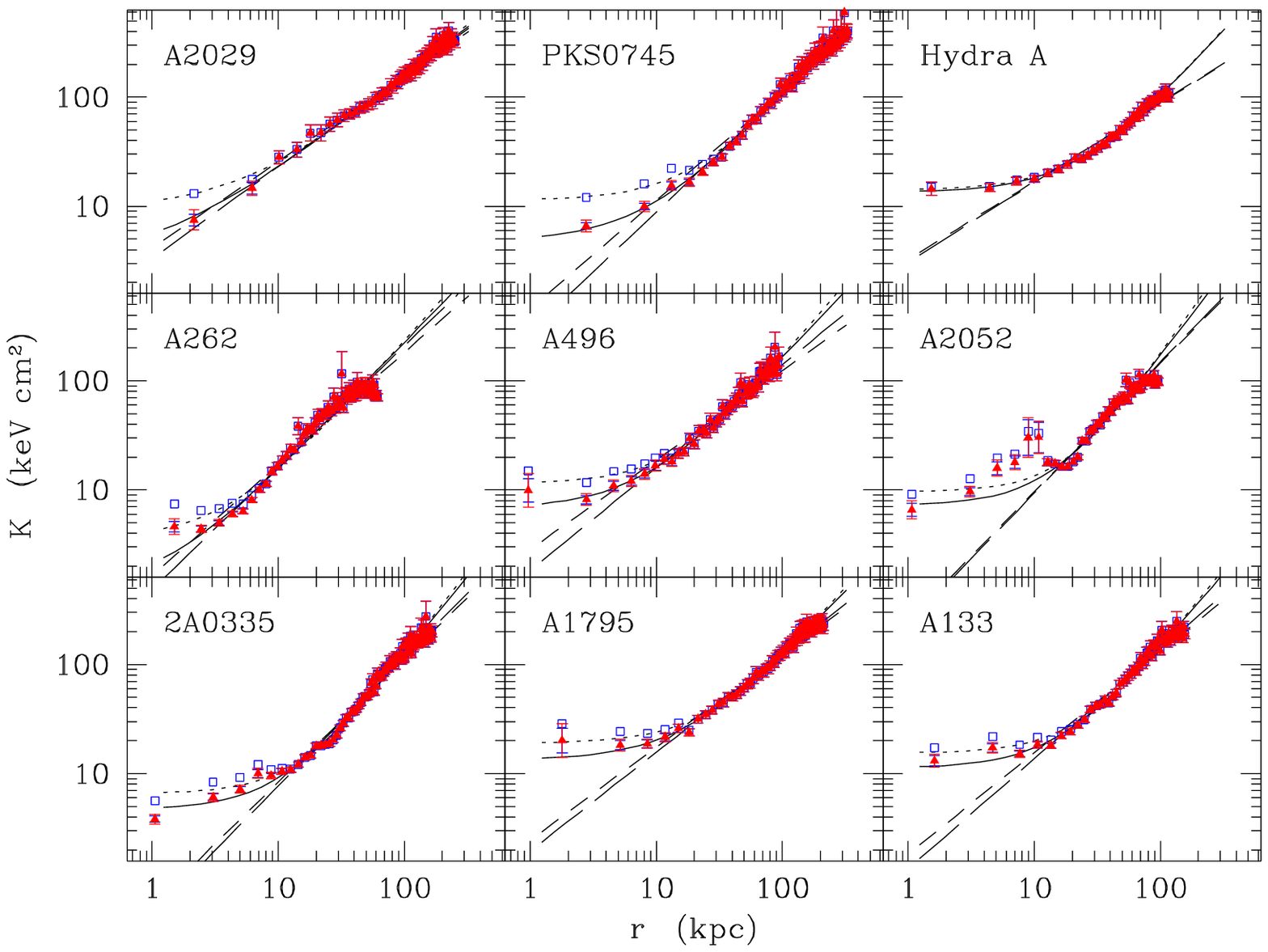}
\caption{Entropy profiles, where $K=kTn_e^{-2/3}$. 
The solid (red) triangles are the profiles derived assuming a temperature gradient within the central temperature bin (Method 1).  Circles show profiles derived assuming a constant temperature in the central temperature bin (Method 2).  The solid and dotted lines are fits to the Method 1 and Method 2 profiles, respectively, using a power law in radius plus a constant entropy level;  the short-dashed and long-dashed are fits to the Method 1 and Method 2 profiles, respectively, using a pure power-law model that falls to zero entropy at $r=0$. 
\label{fig:gallery} }
\end{figure*}

\begin{figure*}
\includegraphics[width=7in,angle=0, trim = 0.7in 1.5in 0.65in 1.6in, clip ]{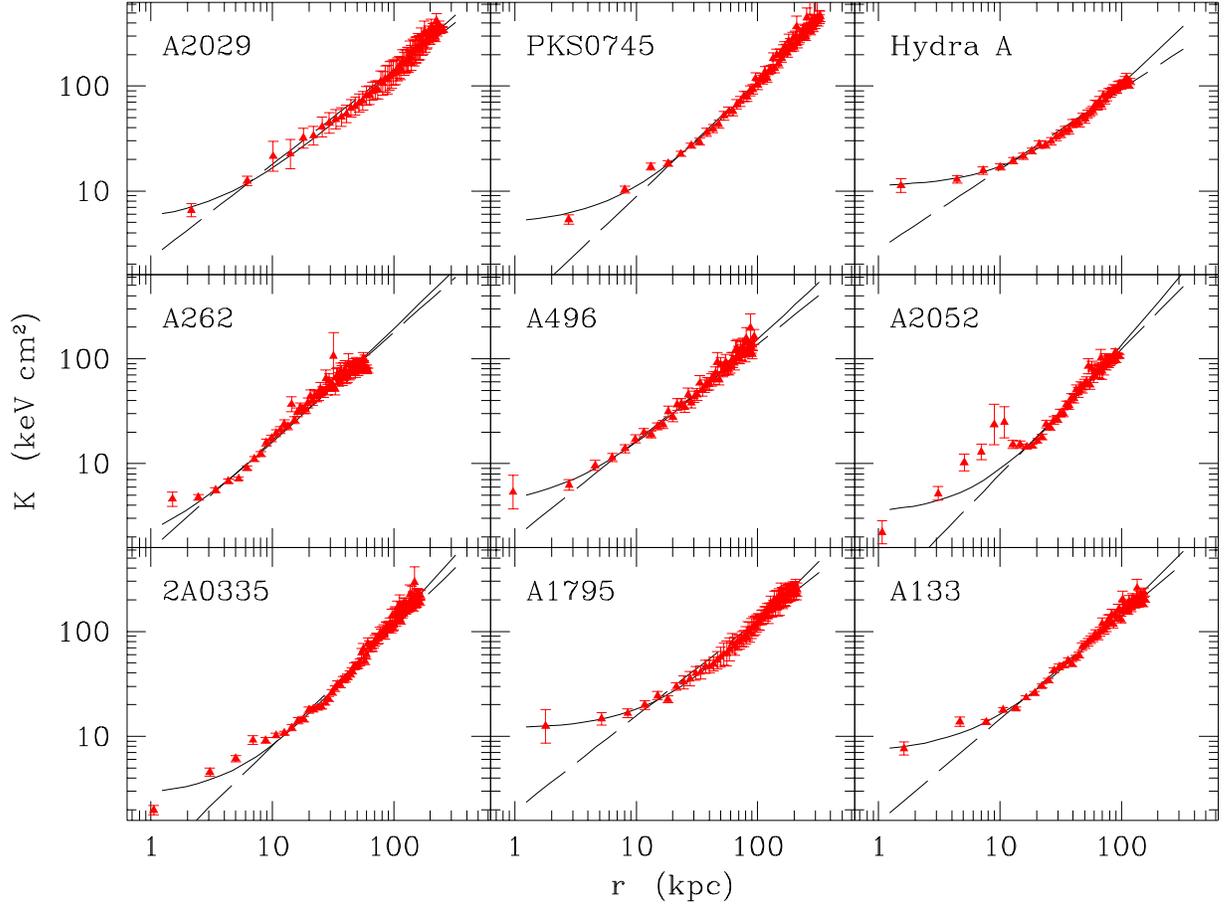}
\caption{Entropy profiles, where $K=kTn_e^{-2/3}$. 
The solid (red) triangles are the entropy profiles where the temperature was 
derived by fitting a simple power 
law ($T \propto r^\alpha$) to the deprojected temperatures (Method 3).  The solid 
lines is a fit to the Method 3 profiles using a power law in radius plus a constant entropy level;  the long-dashed line is a fit to the Method 3 profile 
using a pure power-law model that falls to zero entropy at $r=0$.
\label{fig:gallery_dpt} }
\end{figure*}

Our entropy-profile fitting revealed a remarkable uniformity among the clusters in our sample, which can be seen in Figures~\ref{fig:gallery} and \ref{fig:gallery_dpt}, despite the variety of structures that can be seen in the cluster morphology.
All of them have approximately the same entropy normalization at $\sim 100$~kpc:  $K_{100} \approx 150\pm40$ keV cm$^2$, $K_{100} \approx 150\pm50$ keV cm$^2$, and $K_{100} \approx 140\pm30$ keV cm$^2$ for methods 1, 2, and 3 respectively.  The best fits are generally obtained with a non-zero value for the entropy pedestal:  $K_0 \approx 7\pm4$~keV cm$^2$,  $K_0 \approx 11\pm5$~keV cm$^2$, and $K_0 \approx 6\pm4$~keV cm$^2$ for 
methods 1, 2, and 3 respectively.  They also have similar power-law slopes in the 10-100 kpc range:  $\alpha = 1.2\pm0.2$ for fits with $K_0 \neq 0$, and $\alpha = 1.0\pm0.2$ for fits with $K_0 = 0$.  
Only Abell~2029 has an entropy profile marginally consistent with a vanishing entropy value at $r=0$, having $K_0 = 3.0\pm2.1$ (for 5$\arcsec$ bins) when Method~1 is used for temperature interpolation.  These results are insensitive to the binning procedure, as one can see by comparing Table~\ref{tab:fit-5arcsec} with Table~\ref{tab:fit-2.5arcsec}. The main effect of using a deprojected temperature
profile instead of a projected temperature profile is generally (but not always) a slightly lower central 
entropy quantity $K_0$ (compare Table~\ref{tab:fit-2.5arcsec} with
Table~\ref{tab:fit-2.5arcsec-tdeproj}.) The largest uncertainties come from the statistical uncertainty of the gas temperature and
from the treatment of the temperature profiles in the central bin.

Our results on central entropy values ($K_0$) pertain to the component that fills the majority of the volume in the central bin.  In many cases, there is clearly gas of very low entropy ($K \sim 10^{-5} \, {\rm keV \, cm^2}$ near the center of the cluster in the form of H$\alpha$-emitting nebular filaments.  Another
example of cool, multiphase gas can be seen in the central 10 kpc of M87. The X-ray filaments along the radio source consist of gas at $\sim 1$~keV surrounded by a 2 keV cluster atmosphere \citep[e.g.,][]{Sparks2004}, but those filaments do not constitute a large fraction of the volume in the 10 kpc sphere surrounding M87. 

In order to evaluate how the presence of a cool component affects the entropy derived for the hot volume-filling component, one can consider the emission from a two-component plasma with temperatures $T_h$ and $T_c$ and normalizations $N_h$ and $N_c$ for the hot and cool components, respectively.  If the two components are in pressure equilibrium, then the electron density of the hot component is $propto N_h^{1/2} * [1+(N_c/N_h)(T_c/T_h)^2]^{1/2}$.  In the three clusters for which there is some evidence for a second temperature component, that component is present only within the central bin and has $N_c \sim 0.5 N_h$ and $T_c \sim 0.5 T_h$.  Such a cool component comprises only $\sim 10$\% of the X-ray emitting gas mass, and failing to account for it leads to a derived value of electron density that is only a few percent different from the actual value, if $N_h$ is properly measured.
A larger uncertainty arises in the temperature measurement, because the best-fitting temperature in a single-temperature model will be a weighted mean of $T_c$ and $T_h$ that depends on the spectral band of the fit.  Comparing our single-temperature fits with our two-temperature fits shows that $T_h$ is generally $\sim 50$\% higher than the best-fitting single temperature.  In addition, if a cool component is present, then the normalizations derived from single-temperature models are overestimates of the true normalization of the hot component in the central bin.  In the most extreme case, in which $\sim 50$ of the counts come from the cool component, the electron density inferred from single-temperature models would be $\sim 40$\% larger than the actual value in the hot phase.  Combining all these effects would mean that the central entropy of the hot component could be as much as $\sim 80$\% larger than what we estimate from our single temperature models, if a second component of cooler gas is indeed affecting our temperature measurements, and the entropy of a cool component with $T_c \sim 0.5 T_h$ would be $\sim 50$\% of the single-temperature entropy estimate.

Another source of uncertainty in entropy values derived under the assumption of a single-component plasma crops up in cases where radio plasma creates X-ray cavities.  In such cases, the volume of the X-ray emitting plasma is smaller than assumed, meaning that the electron density is underestimated and the entropy is overestimated.  However, this effect is not large compared with other sources of uncertainty.  If the radio plasma displaces as much as 25\% of the gas in a given annulus, then the true electron density would be $(0.75)^{-1/2} \sim 1.15$ of what was estimated, and the actual entropy would be $(0.75)^{1/3}\sim0.91$ times the estimated value. 

In summary, the entropy profiles we derive here are for the X-ray emitting component that fills the majority of the volume at each radius.  Because the temperature structure of the central bin is difficult to establish, the systematic uncertainties in our central entropy measurements are approximately a factor of two.  In three of our clusters, spectral deprojection suggests that there may be a second temperature component in the central bin at roughly half the temperature of the volume-filling component.  Single-temperature modeling of the central bin could therefore be underestimating the entropy of the volume-filling component.  Our finding that the entropy of the volume-filling component approaches a minimum value $\sim 10 \, {\rm keV \, cm^2}$ at small radii is therefore robust to the presence of a cooler component.  The mass of gas in a cooler X-ray emitting component must be considerably less than that in the hotter component.  Otherwise, it would emit very bright soft X-ray emission lines that would show up in our spectra as a pronounced soft excess. High-resolution spectroscopy of the cores of many of these clusters also severely limits the amount of gas cooler than about $1/2-1/3$ the virial temperature 
\citep{2003ApJ...590..207P}.  In the future, it will be fruitful to compare these results with a parallel analysis of the predicted X-ray emission of simulated 3D clusters with realistic temperature profiles and emission structures.  We are pursuing that work separately.

\subsection{Individual Clusters \label{sec:literature}}

The following paragraphs discuss the individual clusters in our sample.  In each case, we give the radio and
optical emission line characteristics of the cluster, which can affect the X-ray morphology.  Then we discuss how our results on the individual clusters compare with previously published {\em Chandra} analyses.  We also compare our results with published {\em XMM-Newton} results.  All clusters in the sample have a central radio source in a single central
bright galaxy; all but one (Abell 2029) has an optical emission line nebula in the brightest central galaxy. 
We also note that all but one of the clusters  in our
sample with H$\alpha$ emission also have had confirmed detections of vibrationally-excited molecular hydrogen at 2 microns \citep{Edge2002,Donahue2000,Jaffe2001,Falcke1998,JaffeBremer97}, except for 2A0335+096, which has a CO detection by \citet{Edge2001}, and except for
Abell 133, which hasn't been observed for these lines recently.

For the most part, our results on the temperature and metallicity profiles of these clusters seem relatively robust to calibration and telescope differences.  The main discrepancies we find stem from updates to the {\em Chandra} calibration that have reduced the prominence of second temperature components or anomalous absorbing column densities suggested or 
required by the data in previous analyses. For a similar reason, single temperatures derived from fits that included 
all or part of the 0.3-0.7 keV bandpass in those analyses of {\em Chandra} data were occasionally different from what we are now obtaining.  

Our results are in rough agreement with recently published {\em XMM-Newton} temperature profiles for individual clusters. 
{\em XMM-Newton} observations of six of the clusters in our sample were analyzed by \citet{2004A&A...413..415K} and \citet{2004A&A...420..135T}:  2A~0335+096, Abell 262, Abell 2052, Hydra A, Abell 496, and Abell 1795 for the first paper
and 2A~0335+096, Abell 262, and Abell 2052 for the second. For single-temperature projected fits, we
find that the Chandra temperatures are consistently about 0.5 keV higher than XMM temperatures outside $0.5\arcmin$. For
Abell~1795, the  discrepancy is larger, $\sim+1$ keV, over the entire region. For single temperature deprojected fits, we have no
discrepancy with \citet{2004A&A...420..135T} (Abell 262, Abell 2052, and 2A0335+096), except for
the radii of $0.5-1\arcmin$ for 2A0335+096, where we obtain a temperature about 0.4 keV higher. 
For the 2T deprojected fits where a cool component is posited for the center of the cluster, we 
agree with \citet{2004A&A...420..135T}, but for a $\sim+0.2$ keV discrepancy in Abell 262's $0.5-1.0\arcmin$ annulus.
We shall show in the following discussion of individual clusters 
that our analyses rarely disagrees with that of other Chandra observers, so this discrepancy is 
part of a systematic difference between Chandra and XMM that appears to have reduced over time. 
The metallicities we find agree with those found by \citet{2004A&A...420..135T} for all three clusters in their sample.  

The slope and normalization of our entropy measurements are consistent with recent measurements from XMM.   \citet{2005A&A...433..101P} derive and plot entropy profiles from XMM data 
for four clusters in our sample: 2A0335+096, A262, A2052, and A1795. After converting for different Hubble constants, and assuming that their outermost point is coincident with $R_{out}$ listed in their Table 1, we find excellent agreement between the estimated entropy values plotted in their Figure 5 with our entropy measurements for those clusters.  Encouragingly, their entropy profiles at large radii interpolate in to the profiles we measure with very similar slope, and the  
entropy measurements at overlapping radii agree in their normalization.  Our Chandra profiles have higher resolution and 
but span a smaller range of radii. (Abell 496 is also in their sample, but we could not unambiguously 
identify its points in their entropy plot.) 

The fact that our {\em Chandra} results generally agree with the \citet{2004A&A...420..135T} analysis, and the entropy profiles of  \citet{2005A&A...433..101P}, and yet have a nearly constant offset in
temperature with respect to \citet{2004A&A...413..415K}, suggests that the XMM results do not agree with each other.  The discrepancy may 
arise from using different energy ranges in the fitting (\citet{2004A&A...413..415K} fit 0.2-10.0 keV for
every spectrum, while \citet{2004A&A...420..135T} fit 0.3-8.0 keV for the MOS detectors and 0.7-7.0 keV for the pn detector, similar to the energy range we use to fit Chandra data (0.7-7.0 keV).  
Furthermore, \citet{2004A&A...420..135T} use
a more recent version of the XMM calibration and software (SAS 5.3.0 vs SAS 5.3.3). The main difference
between these versions is a new tool (especget) for the extraction of spectra and the computation of 
responses (RMFs and ARFs). Our consistency with later XMM and Chandra results suggests that the
the calibrations of the two missions are leading to increasingly consistent results, which is encouraging news.

We discuss specific comparisons to published analyses of Chandra observations below.

\subsubsection{2A 0335+096}\label{sec:2a}

The cluster 2A0335+096 is poor but popular. \citet{SBO1995} presented VLA images
showing that this cluster contains a double-lobed source with a nucleus. Each lobe lies about $12\arcsec$ on either side of the nucleus of the central galaxy. That galaxy has a spectacular, filamentary emission-line system with a bar and filaments in the central 20 kpc \citep{RomanishinHintzen88}. The X-ray images show two prominent cavities along with filamentary structures that are not correlated with the radio or optical line emission \citep{2003ApJ...596..190M}. The structure of the radio source suggests that it was produced by multiple outbursts.

{\em Chandra} data on this cluster has previously been analyzed by two groups, \citet{2003PASJ...55..585K} and 
\citet{2003ApJ...596..190M}. Their results disagree somewhat with each other. 
Our results agree more with the latter, who may have used a later CALDB version than the
former group.  

Our comparison with the \cite{2003PASJ...55..585K} analysis shows that our projected 1-T fit agrees in the region $0 < r < 20$ kpc, with temperatures between $1.65-1.76$ keV. 
However, the temperature gradient we derive rises more quickly over the $20 < r < 120$ kpc range. At $r \approx 40$ kpc we find a best-fit temperature value $T = 3.0\pm0.1$ keV while \cite{2003PASJ...55..585K} find $T = 1.9\pm^{0.1}_{0.05}$ keV.  At $r \approx 
54- 70$ kpc our best-fit temperature is $3.8\pm0.15$ keV while \cite{2003PASJ...55..585K} obtain $2.5\pm0.1$ keV.
Our overall temperature fits are also more consistent with the ASCA temperature of $2.86\pm0.02$ keV \citep{Horner2001}.  
The difference between our best-fit temperatures and those of  \citet{2003PASJ...55..585K} arises from  the version of the CXC Calibration Database (CALDB) used to calibrate the data and the energy range used for spectral fitting.   We used CALDB 3.0 in our work, while \citet{2003PASJ...55..585K} used CALDB 2.9.  Also, we fit the spectral range $0.7-7.0$ keV, whereas \citet{2003PASJ...55..585K} fit the $0.5-10$ keV range.  In our experience, the earliest results from {\em Chandra} for the coolest clusters are the most affected by improvements in the calibration. Also, including bins at the highest energies without many counts in them can lead to unreliable results.

We also compared our projected and deprojected one-temperature MEKAL fit results with those of  \citet{2003ApJ...596..190M} who used the \citet{2003PASJ...55..585K} data. They do not report the version of CALDB that they used.  The temperature gradient we measure has a similar slope to the one they obtained.   Over the region of 0-240\arcsec, \cite{2003ApJ...596..190M} their best-fit temperature rises from 1.8 keV to 4.2 keV with a flattening at $100\arcsec < R < 200\arcsec$.  The metallicity and $\nh$ profiles are also similar in value and behavior.
A second temperature component in the central region marginally improved our deprojected fit (Table~\ref{tab:table2Tdeproj}), 
but the lower limits on the normalization 
of this component and the modest improvement of $\chi^2$ are not indicative of a secure detection, particularly since
the uncertainties of the calibration are probably highest at the energies of interest here.

\subsubsection{Abell 133}
The center of Abell 133 hosts an impressive radio relic source spanning 55 kpc. \citet{Slee2001} conclude from $4\arcsec$ resolution VLA observations that the central galaxy is not the current source of the relic but may have been where the radio source began. ROSAT X-ray observations show that the
radio source is clearly interacting with the ICM \citep{Rizza2000}.   The central galaxy also has a compact, low-ionization emission-line source \citep{HCW85}.

\citet{2002ApJ...575..764F} report {\em Chandra} observations of Abell 133.   We find statistical agreement with their projected temperature, metallicity, and $\nh$ profiles for the fits in which \citet{2002ApJ...575..764F} restrict the spectral fitting to photons with energies greater than 0.9 keV.
However, their fits for data which include the less energetic photons disagree with ours.  This finding is in accord with the general trend that early {\em Chandra} spectral fitting results involving soft X-ray data 
are probably not reliable because the low-energy calibration for {\em Chandra}  was not well characterized until at least 2004. Hence, the energy range chosen for any given fit can have a significant impact on the best-fit temperatures, absorption column $\nh$, and metallicity values. The fact that the fits of \citet{2002ApJ...575..764F}  that were restricted to higher energies agree with ours supports our suspicion that the {\em Chandra} calibration of the soft X-ray bandpass has improved with time.

\subsubsection{Abell 262}\label{sec:a262}
Abell 262 hosts the weakest radio source of our sample, the doubled-lobed source B2 0149+35 \citep{Parma1986, Fanti1986}, but it also has an impressive emission line nebula \citep{Plana1998}. 
\citet{2004ApJ...612..817B} find that the radio source is anti-correlated with the X-ray emission in their
{\em Chandra} observations and but it is correlated with optical ([N~II]) emission.

In order to have enough bins to fit a 2-parameter power law to the deprojected temperature profile, 
the spectral fits reported for A262 have 10,000 counts per spectrum instead of 20,000.
Our derived temperatures are consistent with the analyses of the Chandra data in \citet{2004ApJ...612..817B}.  Our projected one-temperature fits with $\nh$ fixed to the Galactic value did not describe the inner annulus between
$0$ and $0.35\arcmin$ ($<7$ kpc)  well, with a reduced ${\chi^2}$ of 2.44. 
But, as \cite{2004ApJ...612..817B} also found, adding a second temperature component to the projected data improved our fit with a reduced ${\chi^2}$ of 1.06 for 107 degrees of freedom. We also find good general agreement between their mean temperatures and metallicity and the average temperature and metallicity of our radial fits.   

Fits to deprojected spectra are typically better than those for inner projected spectra, suggesting that departures from single-temperature emission in the projected spectra stem primarily from a steep radial temperature gradient, which leads to superpositions of multi-temperature plasma along the line of sight. This trend was also true for Abell 262.
In the central bin for Abell 262 (i.e., $<0.35\arcmin$), adding a second temperature component to the deprojected model 
only marginally improved the fits to the projected spectra (Table~\ref{tab:table2Tdeproj}).   In these deprojected spectral fits,  a single-temperature model in each radial shell gave a reduced $\chi^2 = 1.28$ for 923 degrees of freedom, while a two-temperature plasma model in the inner two shells give $\chi^2 = 1.22$ for 921 degrees of freedom.

\subsubsection{Abell 496}\label{sec:a496}

A compact radio source (smaller than $\sim 1.5\arcsec$ based on the beam size reported) inhabits the central galaxy in Abell 496 \citep{ODea1995}. The central galaxy is
also the locale for a bright emission-line nebula \citep[e.g. ][]{HBvM1989}. 
Since the observation of Abell 496 had only 60,000 total counts, we first divided this cluster
into three regions. When our spectral analysis of these three regions turned out to be well-behaved,
we expanded the analysis to six regions of 10,000 counts each in order to fit a 2-parameter power-law temperature
profile, and it is this result which is reported in our spectral fits.

\citet{2003ApJ...583L..13D} present the original {\em Chandra} analysis of these data.
Comparison of our projected one-temperature fits with those from their analysis  
show good agreement across the region $0-150$ kpc. Our temperature and
metallicity  profiles are also consistent with the results of deprojected, 
1-T fits to the {\em XMM-Newton} data \citep{2001A&A...379..107T}.

\subsubsection{Abell 1795}\label{sec:a1795}

The central galaxy of Abell 1795 contains a famous extended filament of optical line emission, mapped using long-slit spectroscopy by \citet{Cowie83}. The central bright galaxy also hosts a compact radio source 4C 26.42 \citep{1967MNRAS.135..231C}.  {\em Chandra} observations by \citet{2001MNRAS.321L..33F} revealed a $40\arcsec$ X-ray filament that substantially overlaps the optical emission-line filament.  

The results we report on this cluster agree well with previous X-ray results.  Our projected and deprojected one-temperature and two-temperature fits are in excellent agreement with the
{\em Chandra} data analysis by \cite{2002MNRAS.331..635E}.
They also agree with the projected one-temperature, $\nh$, and metallicity fits from {\em XMM-Newton} analysis of Abell 1795 by \cite{2001A&A...365L..87T}. 

\subsubsection{Abell 2029}\label{sec:a2029}

Abell 2029 occupies a unique niche in our sample because it alone has no trace 
of an optical emission line nebula, but it is still a luminous radio source. Its X-ray image
is remarkably smooth, exhibiting little of the structure seen in some of its sister cool core clusters.
Therefore it has been a textbook example for the quest to determine the self-interaction
cross-section of dark matter by fitting dark matter potentials to the enclosed
mass inferred from X-ray data \cite[e.g.,][]{2003ApJ...586..135L}.

The work of \citet{2002ApJ...573L..13L,2003ApJ...586..135L} on {\em Chandra} observations of Abell 2029 adopts the APEC spectral model within XSPEC and fixes N$_H$ to the Galactic value. Our  projected and deprojected temperature fits, along with our metallicities, agree with that group's analysis.
Our temperature fits confirm the flattening of the temperature profile inside of $18\arcsec$ found by  \citet{2003ApJ...586..135L}.

\subsubsection{Abell 2052}\label{sec:a2052}

The central galaxy of Abell 2052 hosts the radio galaxy 3C 317. \citet{Venturi2004}  found a parsec-scale bipolar radio source with a radiative age of 170 years in VLBA observations of this galaxy and suggested that this source is a restarted radio galaxy. The kpc-scale appearance of the source is that of an amorphous halo with a bright core. The {\em Chandra} X-ray emission shows two bubbles in the ICM \citep{Blanton2003, 2001ApJ...558L..15B}. This galaxy also is the home of a bright emission-line nebula \citep[e.g. ][]{HBvM1989}.

Our projected temperature profiles for fixed $N_{\rm H}$ agree with the {\em Chandra} analysis of \cite{2003ApJ...585..227B}.  We also find that our best-fit metallicities track theirs, increasing to  0.75 solar at  $r \approx 30$, then falling to values around 0.45
solar at larger radii.  Our deprojected temperature and metallicity profiles also agree with theirs.  In particular, we also see a notable 
increase in the deprojected metallicity at $r \approx 30$ kpc.

We also obtained a minimal but likely insignificant improvement in $\chi^2$ when we included a second temperature
component in the innermost bin for the deprojection analysis (Table~\ref{tab:table2Tdeproj}). 
The lower limit on the normalization of this component is quite low, which suggests
that the detection should be treated as an upper limit for the purposes of this paper.

\subsubsection{Hydra A}\label{sec:hydra}

Hydra A was one of the first clusters to be observed by Chandra, and it was
the cluster that sparked our interest in doing this study. Many people have used the
Hydra A data as comparison for their theoretical predictions; we wanted to
provide a larger sample of clusters with similar published measurements.

Our deprojected single-temperature fits agree with the {\em Chandra} results for Hydra A of \cite{2000ApJ...534L.135M}. We also see the temperature jump at $r \approx 70$ kpc with a subsequent decrease in temperature at $r \approx 100$ kpc.  Likewise, comparing our results with those of \cite{2001ApJ...557..546D} for projected 1-T and
2-T MEKAL fits show similar agreement. Our best-fit values for N$_H$, metallicity, and
temperature are consistent with the values in that paper. However, we do not find an improvement in ${\chi^2}$ for the innermost annulus, $0\arcsec < r < 20\arcsec$, when we allow for a second spectral temperature component. This finding is consistent with our results on second temperature components in the cores of other clusters in our sample owing to the changing calibration in the soft energy band on Chandra's ACIS-S detector. The entropy profile from \citet{2001ApJ...557..546D} is completely 
consistent in normalization and shape with the one we present here.

\subsubsection{PKS 0745-191}\label{sec:pks}

PKS0745-191, at $z=0.1028$, is the most distant cluster in our sample. It hosts a
powerful radio source and a luminous emission line nebula.
The results of \cite{2002ApJ...580..763H} on {\em Chandra} data for PKS 0745-191 are consistent with those we obtain from our projected one-temperature spectral fits. The metallicity profiles are also
similar in that they remain nearly constant with a value of $\approx$0.45 solar. The projected two-temperature fits are not relevant here because we did not find a second temperature necessary for any annulus.  Comparison with the results of \cite{2003A&A...407...41C} for {\em XMM-Newton} data for PKS 0745-191 yields consistency between each group's results. For projected and deprojected one-temperature spectral fits with fixed N$_H$, we find a temperature profile similar to \cite{2003A&A...407...41C}.  
Our metallicity profiles are also consistent in values and trend with radius.

\section{Discussion}\label{sec:discussion}

The main new result emerging from this {\em Chandra} study of the core entropy profiles in cooling-flow clusters is that the profiles are quite similar, with several interesting features in common.  Despite the fact that these clusters range in temperature from 2.2 keV to 7.4 keV, their core entropy profiles all have a similar normalization ($\approx 150-160 \, {\rm keV \, cm^2}$) at a radius of 100 kpc, they all have similar power-law slopes within that radius, and the profiles generally tend to flatten to a constant value $\approx 6-10$~keV cm$^2$ at the centers of the clusters.  In this section we examine these features more closely and explore their implications.

\subsection{Observations of Non-Zero Central Entropy}\label{sec:non-zero-obs}

The presence of a non-zero central entropy pedestal in {\em Chandra} observations of Hydra~A was noted by \citet{2001ApJ...557..546D}.  However, it has not generally been recognized as a common feature in cooling-flow clusters.  For example, \citet{2005A&A...433..101P} analyzed the entropy profiles of thirteen cooling-flow clusters observed with {\em XMM-Newton}, finding that they were adequately fit by a pure power law with $\alpha \approx 0.95$ without the need for a central entropy pedestal.  However, the lower spatial resolution of {\em XMM-Newton} ($4\arcsec$) makes it harder to resolve the central $\sim 10$~kpc where the entropy pedestal dominates the entropy profile.  Other {\em Chandra} studies suggest that non-zero central entropy might not be universal in cooling-flow clusters.  One possible counterexample from our own study is Abell~2029.  Another possible counterexample is Abell~478 \citep{2003ApJ...587..619S, Sanderson:2004yz}.  Also, there are group-scale objects which seem to have entropy profiles that tend to zero entropy at the center \citep{2003ApJ...598..250S, 2005ApJ...622..187M}.

\begin{figure*}
\includegraphics[width=7in,angle=0]{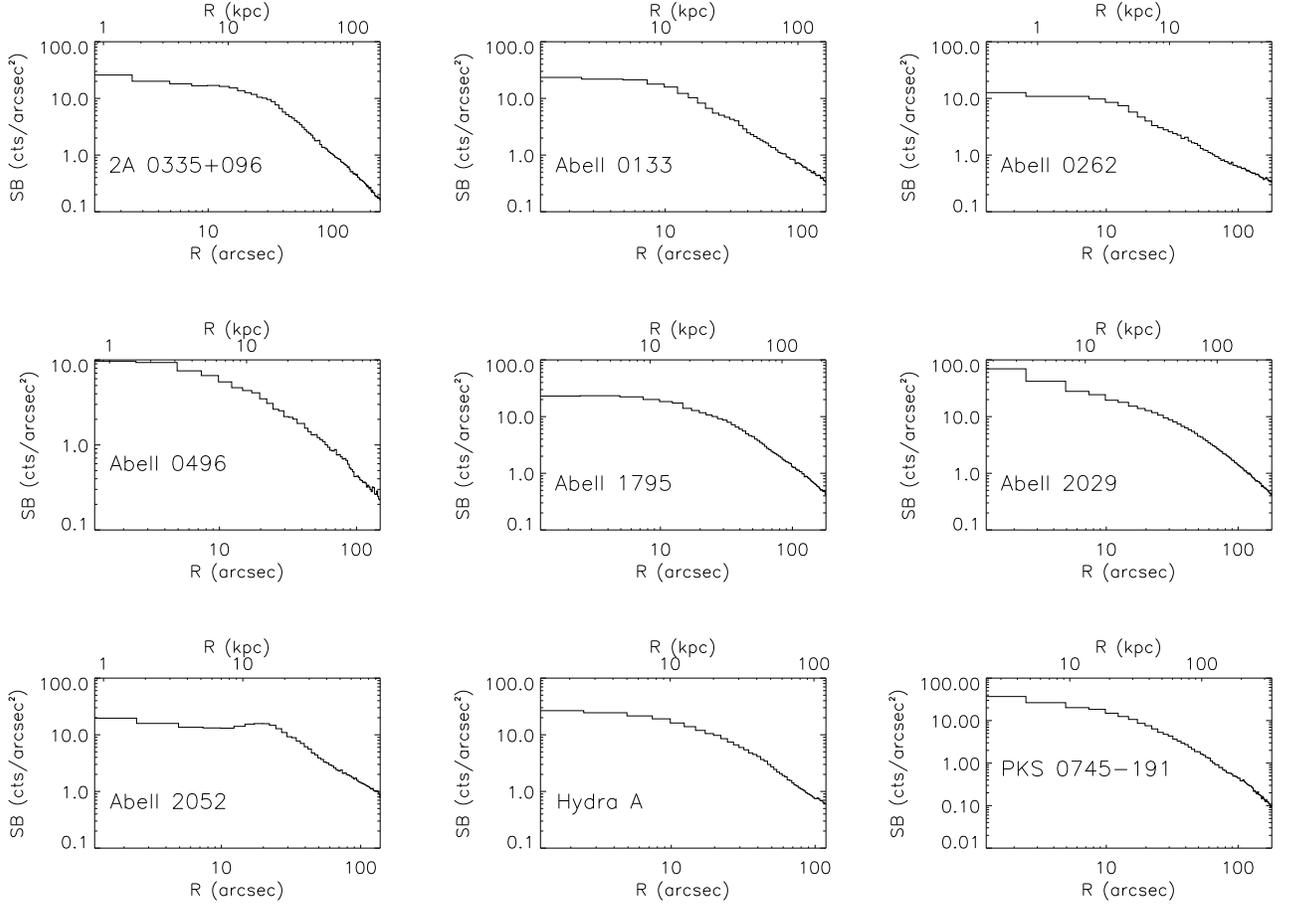}
\caption{Background-subtracted X-ray 
surface brightness profiles, in units of X-ray counts per 
square arcsecond, $0.5-2.0$ keV, where the annuli
are $2.5\arcsec$ wide.  \label{fig:SB_gallery} }
\end{figure*}

\begin{figure}
\plotone{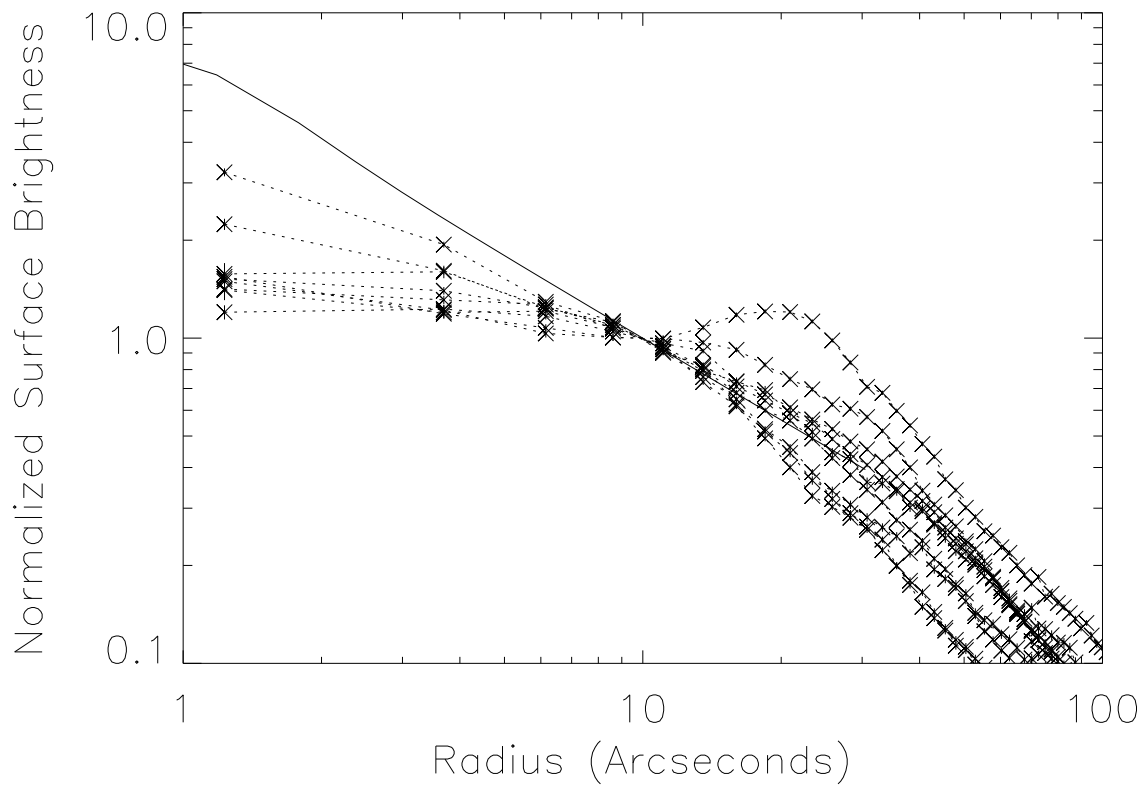}
\caption{The solid line is the predicted surface brightness model for bremstrahlung radiation from
a plasma with $n \sim r^{-1}$ and $T\sim r^{1/3}$. The dotted lines are the surface brightness 
profiles for all 9 clusters in this paper. All surface brightness profiles have been normalized to 1
at $r=10\arcsec$. Note that the predicted model exceeds the observed profiles in the central 10$\arcsec$. The cluster  
with the most peaked surface brightness profile is Abell 2029.
The cluster with the odd hump at $r>10\arcsec$ is Abell 2052. \label{SB_Comparison}}
\end{figure}

Our finding that non-zero central entropy is common, if not universal, among cooling-flow clusters rests on the flattening of the observed surface-brightness profiles within $\sim 10$~kpc of the cluster's center, which typically corresponds to an angular radius of $\sim 10\arcsec$  in these low-redshift clusters (see Figure~\ref{fig:SB_gallery}).  This flattening implies that the electron density profile does not diverge at the center, and therefore that the central entropy tends to some minimum value.  In order to verify that the flattening we observe is not an artifact of the deprojection procedure, we compared our observed surface-brightness profiles with the profile one would expect if the power-law behavior observed at larger radii continued to hold all the way to $r=0$.  For this comparison, we used a power-law core model with $n_e \propto r^{-1}$ and $T(r) \propto r^{1/3}$, so that $K(r) \propto r$.  Assuming a thermal bremstrahlung emissivity ($\propto T^{1/2}$) then yielded a predicted surface-brightness profile $S(\theta) \propto \theta^{-5/6}$. (If we had included line radiation and iron gradients, the central surface-brightness profile would have been even more sharply peaked.)   We then convolved this power-law model with a two-dimensional Gaussian to simulate the effects of a point-spread function.  To be conservative, we simply approximated the Chandra ACIS-S point spread function as a simple Gaussian, $e^{-r^2/\sigma^2}$ with $\sigma=1\arcsec$. This assumed PSF is actually a little broader than the actual one, so we are slightly overestimating the effect of PSF smearing on these profiles. Nevertheless, the surface-brightness profile predicted by this simple power-law model is significantly more peaked than the actual observed surface-brightness profiles, even if we do not include the divergent flux from the central pixel in the calculation.  Figure~\ref{SB_Comparison} shows that comparison with both the observed surface-brightness profiles and the theoretical profile normalized to unity at $r=10\arcsec$.  Therefore, unless the gas temperatures in the cores of these clusters drop much more rapidly than the power-law model we assumed (or indeed more than has been measured in high-resolution
spectroscopy), then they typically have nearly constant central entropy values of $\approx 6-10$ keV cm$^2$.

One could argue that this result is not completely unexpected.  For some time now, X-ray astronomers have been successfully 
fitting ``beta-model'' and even ``double-beta model'' profiles to X-ray surface brightness profiles of clusters 
\citep[e.g., ][]{1998ApJ...496...73M,1999ApJ...517..627M,2000MNRAS.318..715X}. 
The standard ``beta-model" for a cluster's X-ray surface brightness $I(\theta)$ is proportional to $(1+\theta/\theta_c)^{-3\beta+1/2}$, where $\theta_c$ is the projected core radius in appropriate units  \citep{Sarazin1998, 1976A&A....49..137C}.
The  X-ray surface-brightness  data for clusters and groups have consistently 
exhibited evidence for cores in the electron density profile, which is relatively insensitive
to the temperature profile. The key feature of the beta-model is its flat central core. These high resolution Chandra observations 
show that even those clusters with significant peaks in X-ray surface brightness still have cores at $\theta < 10\arcsec$.

\subsection{Implications of Non-Zero Central Entropy}\label{sec:non-zero-imp}

The observed central entropy levels in these nine clusters suggest that the heating mechanism which prevents most of a cooling-flow cluster's core gas from condensing is episodic \citep[see also,][]{2001ApJ...557..546D, 2003MNRAS.338..837K}.  Intracluster gas of entropy $K$ and temperature $T$ that radiates pure thermal bremsstrahlung emission has a cooling time 
\begin{equation}
  t_c \approx 10^8 \, {\rm yr}  \left( \frac {K} {10 \, {\rm keV \, cm^2}} \right)^{3/2}  
  				\left( \frac {kT} {5 \, {\rm keV}} \right)^{-1} \; \; .
\end{equation}
The bulk of the gas currently at the centers of these X-ray clusters therefore will not begin to condense for at least $\sim 100$~Myr. That timescale for the introduction of feedback is consistent with the periodicity of AGN feedback inferred from X-ray studies of the cavities associated with the radio sources at the centers of cooling-flow clusters \citep[e.g.,][]{2004ApJ...607..800B}.  \citet{VoitDonahue2005} show that  an outburst of kinetic power from an AGN at the level of $\sim 10^{45} \, {\rm erg \, s^{-1}}$ can produce such an entropy pedestal through shock heating that raises the core gas of a cooling-flow cluster by a uniform increment of $\sim 10$~keV cm$^2$.

Many clusters have much higher central entropy levels, which correspond to cooling times greater than the age of the universe, meaning that they were never suspected of harboring cooling flows.  One classic example is the Coma cluster.  It is the nearest, richest X-ray cluster, and it has two bright
central cluster galaxies rather than a single dominant one.  The central entropy in the Coma cluster implied by the {\em ROSAT} X-ray observations of \citet{1992A&A...259L..31B} is $340^{+170}_{-80} h_{70}^{-1/3}$ keV cm$^2$.  Such a large central entropy value is difficult to generate purely through feedback, requiring an AGN outburst $\gtrsim 10^{47} \, {\rm erg \, s^{-1}}$ in the heating framework of \citet{VoitDonahue2005}.   We therefore suspect that Coma's core achieved its high entropy level through merger shocks or some other dynamical means.

However, there are other clusters with central entropy levels intermediate between the Coma cluster and the cooling-flow clusters in the present sample.  These came to our attention when we were investigating examples of cooling-flow clusters without evidence for feedback.  Every cluster in the sample we present here has a certain amount of central AGN activity, indicated by the radio power from the central galaxy.  Most of them (all but Abell 2029) also show evidence for active star formation in the form of optical emission-line nebulae.  For the purposes of this discussion, we will call such clusters ``active clusters.''  In order to isolate the relationship between AGN activity and the cooling-flow phenomenon, we used {\em Chandra} to observe two other clusters, Abell 1650 and Abell 2244,  that had been classified as cooling-flow clusters by \citet{Peres1998} but that did not contain measurable radio power from a central source or an optical emission-line nebula.  We will call these two clusters ``passive clusters."

\citet{Donahue2005A} show that the two passive clusters have substantially higher central entropy values than the active clusters.  These central entropy levels, amounting to 30-50~keV cm$^2$, correspond to central cooling times $\sim 1$~Gyr, suggesting that these clusters show no signs of feedback because it is not currently necessary to prevent condensation and may not have been necessary for quite some time in the past.  One possibility is that these clusters were heated by an extraordinarily strong AGN outburst ($\sim 10^{46} \, {\rm erg \, s^{-1}}$) that raised the central entropy level to $\gtrsim 50$~keV cm$^2$ roughly a Gyr or more ago, so that the dynamical traces of that feedback event have now dissipated. Another possibility is that gravitationally driven effects like merger shocks
\citep[e.g., ][]{Buote2005} have kept the central entropy relatively high for the last several Gyr and that these clusters will become more like classic cooling-flow clusters about a Gyr from now \citep[see also][]{2004ApJ...613..811M}.

\subsection{Slopes of Core Entropy Profiles} \label{sec:slopes}

We find a mean power-law slope for core entropy profiles in our sample of $1.2-1.3\pm0.2$ 
when the central entropy is allowed to be non-zero and $0.9-1.0\pm0.2$ for a pure power-law fit that goes to zero at the origin.  Our result for a pure power law agrees with \citet{2005A&A...433..101P}, who found $\alpha \approx 0.95$ in {\em XMM-Newton} observations.  However, the addition of a $\sim 10$~keV~cm$^2$ central entropy pedestal along with a slight steepening of the power-law slope clearly produces a better fit to the {\em Chandra} data, as can be seen in Figure~\ref{fig:gallery}.

In either case, the power-law slope is similar to the power-law index of $\alpha \approx 1.1$ observed for the entropy profiles at larger radii in clusters.  That index seems to be a natural consequence of gravitational structure formation, which naturally produces profiles having $K(r) \propto r^\alpha$ with $\alpha \sim 1.1$ outside the cores of clusters \citep[e.g.,][]{2003A&A...408....1P}.  However, it is not clear why that power-law behavior should continue inside the core, where the physics of cooling and feedback ought to dominate the thermodynamics of intracluster entropy.

\subsection{Entropy Normalization at 100 kpc} \label{sec:100kpc}

The agreement between the entropy levels at radii of 100~kpc in these nine cooling-flow clusters indeed appears to be related to the non-gravitational processes that break self-similarity in the whole population of galaxy clusters.  Gravitational structure formation, acting alone, should produce nearly self-similar clusters whose whose virial radii scale with mass as $r_{\rm vir} \propto M^{1/3}$ and whose temperatures then scale with virial radius as $T \propto r_{\rm vir}^2$.  Because the gas density at a given fraction of the virial radius should be the same for all self-similar clusters, the gas entropy at that fraction of the virial radius should be $K(r/r_{\rm vir}) \propto T$.  For a power-law slope in entropy of $\alpha = 1.1$, the entropy at a given physical radius in self-similar clusters should then scale as $K(r) \propto T^{1-\alpha/2} \propto T^{0.45}$.  Yet we see no such systematic trend in our sample, even though the clusters range over a factor of three in temperature. This finding implies that non-gravitational processes play a role in regulating the entropy level at 100~kpc.

That result should not be surprising, in light of the fact that cooling and feedback are thought to alter the luminosity-temperature relation of clusters through their effects on the core entropy distribution \citep[e.g.,][]{eh91, kaiser91, voit02, 2004MNRAS.348.1078B}.  According to the model of  \citet{vb01}, the entropy levels at the core radii of clusters should be related to the entropy level at which the cooling time of the gas equals a Hubble time, which is
\begin{equation}
  K_c(T) \approx 250 \, {\rm keV \, cm^2} \: \left( \frac {kT} {5 \, {\rm keV}} \right)^{2/3} 
\end{equation}
for pure bremsstrahlung cooling.  In that case, one expects entropy at the core radius of a cluster to be $\propto T^{2/3}$, and observations of entropy at a typical core radius of $0.1 r_{\rm vir}$ do show a general correspondence between $K_c(T)$ and $K(0.1 r_{\rm vir})$ \citep{2003MNRAS.343..331P, voit03}.  

Such a scaling of entropy within the cores of clusters is in accord with the agreement we find in entropy levels at 100~kpc.  If the entropy profiles within the cores of clusters really do scale as $K(r) \propto T^{2/3} (r/r_{\rm vir})^{1.1}$, then we expect $K(100 \, {\rm kpc}) \propto T^{0.12}$ \citep[see also][]{VoitDonahue2005}.  In other words, the observed lack of a trend in temperature in the entropy level at 100~kpc is consistent with the idea that cooling and feedback regulate the entropy levels within the cores of these clusters.

\subsection{A Framework for Cooling and Feedback} \label{sec:framework}

The framework for cooling and feedback presented by \citet{VoitDonahue2005} was motivated by these observational findings and supports the notion that episodic AGN heating is responsible for governing these core entropy profiles.  There we show that the core entropy profile one expects when pure radiative cooling is unopposed by feedback forms a lower bound on the set of entropy profiles we observe here.  That bounding profile was computed by \citet{voit02} by simply allowing radiative losses over a Hubble time to remove entropy from the baseline entropy profile characteristic of gravitational structure formation.  The observed profiles converge to this bounding profile at $\gtrsim 100$~kpc and depart from it at smaller radii, where the bounding profile drops to zero entropy as $r \rightarrow 0$.  Thus, it would appear that some sort of feedback is preventing the observed entropy profiles from converging to the bounding profile.

The magnitude of the central entropy level is an important clue to the feedback mechanism at work.  We show in \citet{VoitDonahue2005} that adding a constant entropy pedestal of 10~keV~cm$^2$ to the bounding profile reproduces the behavior of the observed profiles at small radii.  In the framework we suggest, shock heating by an AGN outflow with constant kinetic power naturally produces a constant entropy pedestal out to $\sim 30$~kpc.  As mentioned above, a kinetic power output $\sim 10^{45} \, {\rm erg \, s^{-1}}$ is required to raise the inner entropy level by $\sim 10 \, {\rm keV \, cm^2}$.  Because that inner entropy level corresponds to a cooling time of $\sim 10^8$~years, episodic AGN outbursts on this timescale are needed to maintain the quasi-steady nature of the entropy profiles of these active clusters.

\section{Summary and Conclusions}\label{sec:summary}

We present temperature gradients, metallicity gradients, flux normalizations, and entropy profiles for a sample of X-ray luminous, nearby clusters of galaxies. Because we selected this particular sample from the {\em Chandra} archive to have large numbers of X-ray counts and singly-peaked surface brightness profiles in order to derive high-quality core entropy profiles, we ended up with a sample of nine classic cooling-flow clusters.  All of them show evidence for feedback, with radio emission from the central galaxy in all cases and central emission-line nebulae in eight out of nine cases, with Abell~2029 being the only exception.  We demonstrate that the entropy profiles in the cores of these clusters are quite similar, with power-law slopes in radius of $\sim1.0-1.3$, flattening to a central entropy plateau $\sim 6-10$ keV cm$^2$ in the central 10 kpc or so. We suggest that the core entropy levels are maintained by periodic feedback from a centrally located AGN, with a duty cycle of about $10^8$ years.  We demonstrate that these non-zero central entropy levels are not an artifact of finite angular resolution of Chandra; in fact, it is the exquisite resolution of Chandra which allows us to unambiguously detect these central entropy levels.

We suggest that classifying clusters of galaxies based on their central entropy levels is a promising way to identify the mechanisms that prevent gas from condensing at their cores.  The ``active clusters" in the sample we present here, with entropy levels $\sim 10 \, {\rm keV  \, cm^2}$, all show evidence for recent feedback, but that feedback has not produced entropy 
inversions in the azimuthally averaged entropy profiles.   We have observed two other clusters with {\em Chandra} that have been classified as cooling-flow clusters, based on the fact that their central cooling time is less than a Hubble time, but that show no evidence for recent feedback.  Those ``passive clusters" turn out to have central entropy levels of $\sim 30-50 \, {\rm keV \, cm^2}$, even though they are as regular in appearance as Abell 2029, which has low central entropy \citep{Donahue2005A}. Such elevated central entropy levels can be produced by an especially strong episode of AGN feedback sometime in the past \citep{VoitDonahue2005}.  It is also plausible that these elevated central entropy levels could be preserved by thermal conduction \citep{Donahue2005A}.  Then there are objects like the Coma cluster, with central entropy levels $\sim 350 \, {\rm keV \, cm^2}$ \citep{1992A&A...259L..31B}, which are difficult to achieve with AGN heating and would therefore seem originate through gravitationally-driven processes like merger shocks.  A larger survey of clusters with a greater variety of central entropy levels are needed to explore these issues.

\hrulefill\hspace{4in}
\acknowledgements

This work was supported through Chandra grants from the Smithsonian Astrophysical Observatory
(GO3-4159X, AR3-4017A, AR5-6016X) and an Astrophysics Theory Program grant (NNG04GI89G).  
This research has made use of data obtained from the Chandra
Data Archive (CDA), which is part of the Chandra X-ray Observatory Science Center,
operated for the National Aeronautics and Space Administration (NASA) 
by the Smithsonian Astrophysical Observatory.
This research has also made use of
data obtained from the High Energy Astrophysics Science Archive
Research Center (HEASARC), provided by NASA's Goddard Space Flight
Center.  This research has made use of the NASA/IPAC Extragalactic
Database (NED) which is operated by the Jet Propulsion Laboratory,
California Institute of Technology, under contract with NASA, and of NASA's Astrophysical
Data System Bibliographic Services.


\bibliography{entropy}


\begin{deluxetable}{lrrrccccccc}
\tablewidth{0pt}
\tabletypesize{\scriptsize}
\tablecaption{Clusters Included in Entropy Library\label{tab:sample}}

\tablehead{
\colhead{Name} & \colhead{ObsID} & \colhead{$\alpha$} & \colhead{$\delta$} & \colhead{Exp. Time} & \colhead{$R_{max}$} & \colhead{$R_{max}$} & \colhead{Net Count Rate} & \colhead{$z$} & \colhead{Log $L_{bol}$} & \colhead{$T_{X}$}\\
\colhead{ } & \colhead{ } & \colhead{[deg.]} & \colhead{[deg.]} & \colhead{[ksec]} & \colhead{[$\arcmin$]} & \colhead{[kpc h$_{70}^{-1}$]} & \colhead{[cts sec$^{-1}$]} & \colhead{ } & \colhead{[$h_{70}^{-2}$ ergs sec$^{-1}$]} & \colhead{[keV]}\\
\colhead{(1)} & \colhead{(2)} & \colhead{(3)} & \colhead{(4)} & \colhead{(5)} & \colhead{(6)} & \colhead{(7)} & \colhead{(8)} & \colhead{(9)} & \colhead{(10)} & \colhead{(11)}\\
}


\startdata
2A0335+096   & 919  & 54.6699  &   9.9668 & 19.98 & 4.0 & 165 & 10.70 & 0.0347 & 44.71 & 2.88\\
A133    & 2203 & 15.6759  & -21.8809 & 35.91 & 2.5 & 161 & 2.36  & 0.0554 & 44.52 & 3.71\\
A262    & 2215 & 28.1948  &  36.1527 & 29.12 & 3.0 & 60  & 2.32  & 0.0163 & 43.73 & 2.17\\
A496    & 3361 & 68.4045  & -13.2608 & 10.13 & 2.5 & 95  & 6.49  & 0.0317 & 44.61 & 3.89\\
A1795    & 493  & 207.2207 &  26.5907 & 19.88 & 3.0 & 216 & 10.44 & 0.0622 & 45.21 & 5.49\\
A2029    & 891  & 227.7248 &   5.7451 & 20.59 & 3.0 & 260 & 11.57 & 0.0761 & 45.49 & 7.38\\
A2052    & 890  & 229.1834 &   7.0211 & 37.23 & 2.3 & 97  & 4.63  & 0.0353 & 44.44 & 2.96\\
Hydra A       & 576  & 139.5241 & -12.0955 & 19.88 & 2.0 & 122 & 5.93  & 0.0522 & 44.79 & 3.54\\
PKS0745-191  & 2427 & 116.8803 & -19.2944 & 18.09 & 3.0 & 340 & 5.96  & 0.1028 & 45.66 & 6.25\\
\enddata
\end{deluxetable}

\begin{deluxetable}{lrrrrrrrrrr}
\tablewidth{0pt}
\tabletypesize{\scriptsize}
\tablecaption{Results for 1-T Projected, $\nh$ Fixed MEKAL Fits\label{tab:1Tprfixed}}
\tablehead{\colhead{Name} & \colhead{R$_{out}$ } & \colhead{$\nh$ } & \colhead{T$_{X}$ } & \colhead{Norm} & \colhead{$\sigma_{Norm}$ } & \colhead{Z} & \colhead{$\chi^2_{red}$ } & \colhead{d.o.f}\\
\colhead{ } & \colhead{kpc} & \colhead{$10^{20}$ cm$^{-2}$} & \colhead{keV} & \colhead{ } & \colhead{ } & \colhead{$Z_{\odot}$ } & \colhead{ } & \colhead{ }\\
\colhead{{(1)}} & \colhead{{(2)}} & \colhead{{(3)}} & \colhead{{(4)}} & \colhead{{(5)}} & \colhead{{(6)}} & \colhead{{(7)}} & \colhead{{(8)}} & \colhead{{(9)}}
}
\startdata
2A0335+096 & 12.44   & 18.11 & 1.70$^{+0.04   }_{-0.04   }$  & 6.91e-03 & 3.00e-04 & 0.57$^{+0.06   }_{-0.05   }$  & 1.71 & 166\\
  & 19.07   & -- & 2.04$^{+0.05   }_{-0.05   }$  & 5.98e-03 & 2.60e-04 & 0.80$^{+0.09   }_{-0.08   }$  & 1.28 & 173\\
  & 25.71   & -- & 2.24$^{+0.05   }_{-0.05   }$  & 6.82e-03 & 2.60e-04 & 0.77$^{+0.08   }_{-0.07   }$  & 1.31 & 183\\
  & 33.17   & -- & 2.64$^{+0.07   }_{-0.07   }$  & 5.78e-03 & 2.20e-04 & 0.93$^{+0.10   }_{-0.09   }$  & 1.48 & 193\\
  & 42.29   & -- & 3.03$^{+0.08   }_{-0.08   }$  & 6.37e-03 & 2.10e-04 & 0.81$^{+0.09   }_{-0.08   }$  & 1.42 & 207\\
  & 54.31   & -- & 3.35$^{+0.11   }_{-0.08   }$  & 6.41e-03 & 2.10e-04 & 0.82$^{+0.09   }_{-0.09   }$  & 1.32 & 218\\
  & 70.48   & -- & 3.85$^{+0.13   }_{-0.12   }$  & 6.80e-03 & 2.00e-04 & 0.74$^{+0.10   }_{-0.08   }$  & 1.40 & 233\\
  & 90.80   & -- & 4.00$^{+0.14   }_{-0.13   }$  & 7.20e-03 & 2.20e-04 & 0.78$^{+0.10   }_{-0.09   }$  & 1.11 & 246\\
  & 165.84  & -- & 4.43$^{+0.12   }_{-0.11   }$  & 1.77e-02 & 3.00e-04 & 0.61$^{+0.06   }_{-0.06   }$  & 1.45 & 346\\
A133 & 23.24   & 1.58  & 2.31$^{+0.07   }_{-0.06   }$  & 2.14e-03 & 1.00e-04 & 1.19$^{+0.14   }_{-0.11   }$  & 1.29 & 171\\
  & 49.07   & -- & 3.00$^{+0.10   }_{-0.10   }$  & 2.58e-03 & 9.00e-05 & 0.92$^{+0.11   }_{-0.10   }$  & 1.21 & 190\\
  & 89.74   & -- & 4.10$^{+0.17   }_{-0.15   }$  & 3.12e-03 & 1.00e-04 & 0.69$^{+0.11   }_{-0.09   }$  & 1.06 & 223\\
  & 161.40  & -- & 4.44$^{+0.19   }_{-0.17   }$  & 4.53e-03 & 1.20e-04 & 0.43$^{+0.08   }_{-0.08   }$  & 0.97 & 274\\
A262 & 6.97    & 5.46  & 1.12$^{+0.02   }_{-0.02   }$  & 1.92e-03 & 1.20e-04 & 0.28$^{+0.04   }_{-0.03   }$  & 2.44 & 109\\
  & 13.74   & -- & 1.66$^{+0.04   }_{-0.04   }$  & 1.27e-03 & 8.00e-05 & 1.12$^{+0.15   }_{-0.13   }$  & 1.79 & 129\\
  & 21.12   & -- & 2.03$^{+0.07   }_{-0.06   }$  & 1.46e-03 & 1.00e-04 & 1.21$^{+0.19   }_{-0.14   }$  & 1.12 & 137\\
  & 30.48   & -- & 2.25$^{+0.08   }_{-0.07   }$  & 1.86e-03 & 9.00e-05 & 1.07$^{+0.14   }_{-0.13   }$  & 1.13 & 154\\
  & 40.64   & -- & 2.49$^{+0.09   }_{-0.09   }$  & 2.29e-03 & 1.10e-04 & 0.86$^{+0.11   }_{-0.11   }$  & 1.01 & 169\\
  & 59.76   & -- & 2.32$^{+0.06   }_{-0.06   }$  & 4.93e-03 & 1.50e-04 & 0.51$^{+0.06   }_{-0.05   }$  & 1.38 & 223\\
A496 & 15.57   & 4.80  & 2.37$^{+0.09   }_{-0.10   }$  & 4.86e-03 & 2.80e-04 & 0.95$^{+0.15   }_{-0.13   }$  & 1.39 & 138\\
  & 28.11   & -- & 3.09$^{+0.14   }_{-0.14   }$  & 5.48e-03 & 2.60e-04 & 0.79$^{+0.14   }_{-0.12   }$  & 0.85 & 147\\
  & 41.78   & -- & 3.82$^{+0.19   }_{-0.19   }$  & 5.42e-03 & 2.60e-04 & 0.95$^{+0.18   }_{-0.15   }$  & 1.04 & 156\\
  & 57.35   & -- & 4.06$^{+0.23   }_{-0.21   }$  & 5.93e-03 & 2.50e-04 & 0.69$^{+0.14   }_{-0.13   }$  & 1.23 & 160\\
  & 78.24   & -- & 4.82$^{+0.31   }_{-0.28   }$  & 6.57e-03 & 2.60e-04 & 0.62$^{+0.16   }_{-0.14   }$  & 1.08 & 172\\
  & 94.95   & -- & 5.11$^{+0.42   }_{-0.37   }$  & 4.92e-03 & 2.40e-04 & 0.50$^{+0.17   }_{-0.18   }$  & 1.19 & 155\\
A1795 & 20.86   & 1.17  & 3.56$^{+0.13   }_{-0.13   }$  & 4.47e-03 & 1.40e-04 & 0.62$^{+0.08   }_{-0.09   }$  & 1.06 & 186\\
  & 33.81   & -- & 4.33$^{+0.19   }_{-0.18   }$  & 4.61e-03 & 1.30e-04 & 0.60$^{+0.10   }_{-0.09   }$  & 1.04 & 197\\
  & 46.04   & -- & 4.66$^{+0.21   }_{-0.20   }$  & 4.51e-03 & 1.40e-04 & 0.72$^{+0.11   }_{-0.11   }$  & 1.11 & 199\\
  & 59.71   & -- & 5.20$^{+0.28   }_{-0.26   }$  & 4.79e-03 & 1.20e-04 & 0.46$^{+0.10   }_{-0.09   }$  & 1.01 & 206\\
  & 74.82   & -- & 5.34$^{+0.29   }_{-0.26   }$  & 4.92e-03 & 1.30e-04 & 0.47$^{+0.10   }_{-0.11   }$  & 0.86 & 210\\
  & 93.52   & -- & 5.80$^{+0.33   }_{-0.30   }$  & 4.96e-03 & 1.30e-04 & 0.42$^{+0.11   }_{-0.09   }$  & 1.03 & 214\\
  & 114.38  & -- & 5.89$^{+0.34   }_{-0.29   }$  & 4.92e-03 & 1.40e-04 & 0.55$^{+0.12   }_{-0.12   }$  & 1.19 & 216\\
  & 139.56  & -- & 6.57$^{+0.42   }_{-0.38   }$  & 5.32e-03 & 1.30e-04 & 0.31$^{+0.11   }_{-0.11   }$  & 1.02 & 226\\
  & 171.22  & -- & 6.39$^{+0.63   }_{-0.53   }$  & 3.85e-03 & 1.60e-04 & 0.50$^{+0.19   }_{-0.16   }$  & 1.11 & 231\\
  & 215.82  & -- & 6.47$^{+0.41   }_{-0.38   }$  & 6.79e-03 & 1.50e-04 & 0.12$^{+0.09   }_{-0.08   }$  & 1.09 & 252\\
A2029 & 9.52    & 3.15  & 4.14$^{+0.28   }_{-0.27   }$  & 1.28e-03 & 8.00e-05 & 1.32$^{+0.30   }_{-0.27   }$  & 1.31 & 124\\
  & 15.58   & -- & 5.94$^{+0.58   }_{-0.48   }$  & 1.72e-03 & 9.00e-05 & 0.65$^{+0.22   }_{-0.21   }$  & 1.21 & 138\\
  & 20.78   & -- & 6.04$^{+0.62   }_{-0.51   }$  & 1.36e-03 & 9.00e-05 & 1.26$^{+0.36   }_{-0.31   }$  & 1.04 & 129\\
  & 34.63   & -- & 7.16$^{+0.43   }_{-0.39   }$  & 4.99e-03 & 1.40e-04 & 0.67$^{+0.14   }_{-0.14   }$  & 1.05 & 225\\
  & 48.48   & -- & 7.07$^{+0.43   }_{-0.39   }$  & 5.02e-03 & 1.40e-04 & 0.59$^{+0.14   }_{-0.13   }$  & 0.97 & 226\\
  & 62.34   & -- & 7.39$^{+0.51   }_{-0.46   }$  & 5.24e-03 & 1.30e-04 & 0.40$^{+0.11   }_{-0.11   }$  & 1.14 & 227\\
  & 77.92   & -- & 7.73$^{+0.51   }_{-0.46   }$  & 5.20e-03 & 1.40e-04 & 0.59$^{+0.14   }_{-0.13   }$  & 0.93 & 226\\
  & 95.24   & -- & 8.56$^{+0.68   }_{-0.58   }$  & 5.23e-03 & 1.50e-04 & 0.49$^{+0.16   }_{-0.14   }$  & 1.12 & 229\\
  & 115.15  & -- & 8.12$^{+0.71   }_{-0.61   }$  & 4.92e-03 & 1.40e-04 & 0.44$^{+0.15   }_{-0.13   }$  & 1.26 & 230\\
  & 138.53  & -- & 8.67$^{+0.68   }_{-0.60   }$  & 5.63e-03 & 1.40e-04 & 0.36$^{+0.13   }_{-0.13   }$  & 1.05 & 235\\
  & 167.10  & -- & 9.19$^{+0.79   }_{-0.66   }$  & 5.71e-03 & 1.50e-04 & 0.38$^{+0.14   }_{-0.14   }$  & 0.98 & 240\\
  & 202.60  & -- & 8.74$^{+0.71   }_{-0.60   }$  & 5.85e-03 & 1.60e-04 & 0.44$^{+0.15   }_{-0.14   }$  & 1.12 & 244\\
  & 259.74  & -- & 9.01$^{+0.65   }_{-0.57   }$  & 7.60e-03 & 1.80e-04 & 0.40$^{+0.14   }_{-0.12   }$  & 1.05 & 275\\
A2052 & 13.90   & 2.85  & 1.67$^{+0.04   }_{-0.04   }$  & 2.42e-03 & 9.00e-05 & 0.49$^{+0.05   }_{-0.05   }$  & 2.59 & 150\\
  & 20.22   & -- & 1.91$^{+0.05   }_{-0.05   }$  & 2.54e-03 & 9.00e-05 & 0.59$^{+0.06   }_{-0.06   }$  & 1.83 & 164\\
  & 26.96   & -- & 2.77$^{+0.08   }_{-0.08   }$  & 2.24e-03 & 8.00e-05 & 0.97$^{+0.10   }_{-0.10   }$  & 1.30 & 181\\
  & 35.38   & -- & 2.93$^{+0.09   }_{-0.09   }$  & 2.55e-03 & 8.00e-05 & 0.68$^{+0.08   }_{-0.07   }$  & 1.17 & 186\\
  & 46.75   & -- & 3.23$^{+0.11   }_{-0.11   }$  & 2.59e-03 & 9.00e-05 & 0.72$^{+0.10   }_{-0.08   }$  & 1.12 & 197\\
  & 59.39   & -- & 3.34$^{+0.12   }_{-0.11   }$  & 2.78e-03 & 8.00e-05 & 0.57$^{+0.08   }_{-0.08   }$  & 1.08 & 206\\
  & 73.71   & -- & 3.51$^{+0.13   }_{-0.12   }$  & 2.94e-03 & 9.00e-05 & 0.50$^{+0.08   }_{-0.07   }$  & 1.09 & 209\\
  & 96.88   & -- & 3.33$^{+0.10   }_{-0.11   }$  & 4.51e-03 & 1.10e-04 & 0.46$^{+0.06   }_{-0.05   }$  & 1.15 & 248\\
Hydra A & 19.55   & 4.84  & 3.06$^{+0.11   }_{-0.10   }$  & 5.50e-03 & 1.50e-04 & 0.49$^{+0.06   }_{-0.06   }$  & 1.12 & 186\\
  & 33.59   & -- & 3.17$^{+0.12   }_{-0.11   }$  & 5.38e-03 & 1.50e-04 & 0.37$^{+0.05   }_{-0.06   }$  & 1.28 & 182\\
  & 49.47   & -- & 3.47$^{+0.14   }_{-0.13   }$  & 5.16e-03 & 1.40e-04 & 0.41$^{+0.07   }_{-0.06   }$  & 1.26 & 191\\
  & 72.69   & -- & 3.59$^{+0.15   }_{-0.14   }$  & 5.38e-03 & 1.50e-04 & 0.35$^{+0.06   }_{-0.06   }$  & 1.17 & 195\\
  & 122.16  & -- & 3.66$^{+0.14   }_{-0.13   }$  & 8.45e-03 & 1.90e-04 & 0.28$^{+0.05   }_{-0.05   }$  & 1.00 & 242\\
PKS0745-191 & 31.75   & 43.39 & 4.08$^{+0.14   }_{-0.13   }$  & 1.17e-02 & 3.00e-04 & 0.50$^{+0.07   }_{-0.07   }$  & 1.24 & 261\\
  & 57.83   & -- & 5.87$^{+0.28   }_{-0.26   }$  & 1.12e-02 & 3.00e-04 & 0.40$^{+0.08   }_{-0.07   }$  & 0.99 & 276\\
  & 91.85   & -- & 6.96$^{+0.37   }_{-0.34   }$  & 1.09e-02 & 2.00e-04 & 0.50$^{+0.09   }_{-0.09   }$  & 1.03 & 286\\
  & 147.42  & -- & 7.47$^{+0.44   }_{-0.39   }$  & 1.20e-02 & 3.00e-04 & 0.37$^{+0.08   }_{-0.08   }$  & 1.14 & 293\\
  & 340.20  & -- & 8.56$^{+0.46   }_{-0.41   }$  & 2.04e-02 & 3.00e-04 & 0.26$^{+0.07   }_{-0.07   }$  & 1.05 & 353\\
\enddata
\end{deluxetable}

\begin{deluxetable}{lrrrrrrrrrr}
\tablewidth{0pt}
\tabletypesize{\scriptsize}
\tablecaption{Results for 1-T Deprojected, $\nh$ Fixed MEKAL Fits\label{tab:table1Tdeproj}}
\tablehead{\colhead{Name} & \colhead{R$_{out}$ } & \colhead{$\nh$ } & \colhead{T$_{X}$ } & \colhead{Norm} & \colhead{$\sigma_{Norm}$ } & \colhead{Z} & \colhead{$\chi^2_{red}$ } & \colhead{d.o.f}\\
\colhead{ } & \colhead{kpc} & \colhead{$10^{20}$ cm$^{-2}$} & \colhead{keV} & \colhead{ } & \colhead{ } & \colhead{$Z_{\odot}$ } & \colhead{ } & \colhead{ }\\
\colhead{{(1)}} & \colhead{{(2)}} & \colhead{{(3)}} & \colhead{{(4)}} & \colhead{{(5)}} & \colhead{{(6)}} & \colhead{{(7)}} & \colhead{{(8)}} & \colhead{{(9)}}
}
\startdata
2A0335+096 & 12.44   & 18.11 & 1.37$^{+0.06   }_{-0.04   }$  & 2.55e-03 & 2.50e-04 & 0.57$^{+0.09   }_{-0.08   }$  & 1.34 & 1970\\
  & 19.07   & -- & 1.69$^{+0.12   }_{-0.12   }$  & 3.62e-03 & 2.70e-04 & -- & -- & --\\
  & 25.71   & -- & 1.97$^{+0.10   }_{-0.08   }$  & 6.00e-03 & 3.50e-04 & 0.87$^{+0.10   }_{-0.09   }$  & -- & --\\
  & 33.17   & -- & 2.23$^{+0.13   }_{-0.11   }$  & 5.15e-03 & 3.00e-04 & -- & -- & --\\
  & 42.29   & -- & 2.74$^{+0.15   }_{-0.14   }$  & 6.36e-03 & 2.60e-04 & 0.86$^{+0.07   }_{-0.07   }$  & -- & --\\
  & 54.31   & -- & 2.95$^{+0.18   }_{-0.16   }$  & 6.51e-03 & 2.50e-04 & -- & -- & --\\
  & 70.48   & -- & 3.63$^{+0.31   }_{-0.30   }$  & 5.77e-03 & 2.30e-04 & -- & -- & --\\
  & 90.80   & -- & 3.69$^{+0.23   }_{-0.22   }$  & 8.58e-03 & 2.30e-04 & 0.65$^{+0.05   }_{-0.05   }$  & -- & --\\
  & 165.84  & -- & 4.45$^{+0.12   }_{-0.11   }$  & 2.50e-02 & 4.00e-04 & -- & -- & --\\
A133 & 23.24   & 1.58  & 2.07$^{+0.07   }_{-0.07   }$  & 1.44e-03 & 7.00e-05 & 1.18$^{+0.11   }_{-0.11   }$  & 1.10 & 860\\
  & 49.07   & -- & 2.72$^{+0.12   }_{-0.12   }$  & 1.99e-03 & 9.00e-05 & -- & -- & --\\
  & 89.74   & -- & 3.69$^{+0.29   }_{-0.26   }$  & 2.56e-03 & 8.00e-05 & 0.55$^{+0.07   }_{-0.06   }$  & -- & --\\
  & 161.40  & -- & 4.53$^{+0.18   }_{-0.17   }$  & 6.34e-03 & 1.40e-04 & -- & -- & --\\
A262 & 6.97    & 5.46  & 0.95$^{+0.02   }_{-0.03   }$  & 1.21e-03 & 1.60e-04 & 0.26$^{+0.06   }_{-0.04   }$  & 1.28 & 923\\
  & 13.74   & -- & 1.46$^{+0.06   }_{-0.05   }$  & 5.94e-04 & 8.50e-05 & 1.59$^{+0.25   }_{-0.21   }$  & -- & --\\
  & 21.12   & -- & 1.90$^{+0.14   }_{-0.12   }$  & 9.12e-04 & 9.00e-05 & -- & -- & --\\
  & 30.48   & -- & 2.02$^{+0.17   }_{-0.14   }$  & 1.04e-03 & 1.17e-04 & -- & -- & --\\
  & 40.64   & -- & 2.65$^{+0.34   }_{-0.30   }$  & 9.75e-04 & 2.39e-04 & 1.86$^{+1.05   }_{-0.39   }$  & -- & --\\
  & 59.76   & -- & 2.35$^{+0.08   }_{-0.06   }$  & 8.46e-03 & 2.80e-04 & 0.53$^{+0.07   }_{-0.05   }$  & -- & --\\
A496 & 15.57   & 4.80  & 2.00$^{+0.14   }_{-0.12   }$  & 2.64e-03 & 3.10e-04 & 0.98$^{+0.30   }_{-0.23   }$  & 1.12 & 928\\
  & 28.11   & -- & 2.54$^{+0.28   }_{-0.25   }$  & 3.95e-03 & 4.40e-04 & 0.69$^{+0.28   }_{-0.21   }$  & -- & --\\
  & 41.78   & -- & 3.55$^{+0.45   }_{-0.42   }$  & 3.90e-03 & 4.90e-04 & 1.19$^{+0.51   }_{-0.40   }$  & -- & --\\
  & 57.35   & -- & 3.61$^{+0.43   }_{-0.37   }$  & 5.85e-03 & 5.50e-04 & 0.76$^{+0.32   }_{-0.25   }$  & -- & --\\
  & 78.24   & -- & 4.37$^{+0.92   }_{-0.65   }$  & 4.91e-03 & 6.40e-04 & 0.80$^{+0.53   }_{-0.40   }$  & -- & --\\
  & 94.95   & -- & 5.19$^{+0.47   }_{-0.41   }$  & 1.20e-02 & 6.00e-04 & 0.48$^{+0.18   }_{-0.18   }$  & -- & --\\
A1795 & 20.86   & 1.17  & 2.84$^{+0.25   }_{-0.23   }$  & 1.96e-03 & 9.00e-05 & 0.69$^{+0.09   }_{-0.09   }$  & 1.06 & 2144\\
  & 33.81   & -- & 4.11$^{+0.66   }_{-0.54   }$  & 2.63e-03 & 1.30e-04 & -- & -- & --\\
  & 46.04   & -- & 3.70$^{+0.55   }_{-0.58   }$  & 3.70e-03 & 1.60e-04 & -- & -- & --\\
  & 59.71   & -- & 5.69$^{+1.64   }_{-1.25   }$  & 3.68e-03 & 1.70e-04 & 0.51$^{+0.09   }_{-0.10   }$  & -- & --\\
  & 74.82   & -- & 4.49$^{+0.88   }_{-0.60   }$  & 4.90e-03 & 2.20e-04 & -- & -- & --\\
  & 93.52   & -- & 6.07$^{+1.42   }_{-1.15   }$  & 4.29e-03 & 2.00e-04 & -- & -- & --\\
  & 114.38  & -- & 4.93$^{+1.06   }_{-0.70   }$  & 4.82e-03 & 2.00e-04 & 0.42$^{+0.07   }_{-0.07   }$  & -- & --\\
  & 139.56  & -- & 6.88$^{+1.05   }_{-0.85   }$  & 7.44e-03 & 2.10e-04 & -- & -- & --\\
  & 171.22  & -- & 5.79$^{+1.62   }_{-1.14   }$  & 4.70e-03 & 2.30e-04 & -- & -- & --\\
  & 215.82  & -- & 6.98$^{+1.20   }_{-0.96   }$  & 7.95e-03 & 2.40e-04 & -- & -- & --\\
A2029 & 9.52    & 3.15  & 2.51$^{+0.35   }_{-0.28   }$  & 5.70e-04 & 4.70e-05 & 0.75$^{+0.08   }_{-0.10   }$  & 1.10 & 2756\\
  & 15.58   & -- & 6.35$^{+2.48   }_{-0.77   }$  & 9.85e-04 & 7.50e-05 & -- & -- & --\\
  & 20.78   & -- & 3.07$^{+1.17   }_{-0.68   }$  & 6.48e-04 & 7.80e-05 & -- & -- & --\\
  & 34.63   & -- & 7.39$^{+1.37   }_{-0.99   }$  & 3.24e-03 & 1.30e-04 & -- & -- & --\\
  & 48.48   & -- & 6.51$^{+1.24   }_{-1.00   }$  & 3.19e-03 & 1.60e-04 & -- & -- & --\\
  & 62.34   & -- & 6.96$^{+1.58   }_{-1.11   }$  & 4.22e-03 & 2.10e-04 & 0.49$^{+0.15   }_{-0.16   }$  & -- & --\\
  & 77.92   & -- & 6.52$^{+1.60   }_{-1.10   }$  & 4.51e-03 & 2.20e-04 & -- & -- & --\\
  & 95.24   & -- & 9.71$^{+3.21   }_{-2.01   }$  & 5.32e-03 & 2.90e-04 & 0.65$^{+0.25   }_{-0.12   }$  & -- & --\\
  & 115.15  & -- & 6.44$^{+2.54   }_{-1.32   }$  & 3.72e-03 & 2.80e-04 & -- & -- & --\\
  & 138.53  & -- & 8.41$^{+2.95   }_{-1.71   }$  & 5.77e-03 & 3.20e-04 & 0.30$^{+0.17   }_{-0.20   }$  & -- & --\\
  & 167.10  & -- & 9.95$^{+3.26   }_{-2.07   }$  & 6.34e-03 & 2.80e-04 & -- & -- & --\\
  & 202.60  & -- & 8.43$^{+1.38   }_{-1.74   }$  & 5.79e-03 & 4.40e-04 & 0.01$^{+0.39   }_{-0.01   }$  & -- & --\\
  & 259.74  & -- & 9.02$^{+0.64   }_{-0.48   }$  & 1.55e-02 & 3.00e-04 & 0.75$^{+0.08   }_{-0.10   }$  & -- & --\\
A2052 & 13.90   & 2.85  & 1.13$^{+0.10   }_{-0.08   }$  & 3.59e-04 & 7.20e-05 & 0.49$^{+0.09   }_{-0.08   }$  & 1.21 & 1545\\
  & 20.22   & -- & 1.40$^{+0.06   }_{-0.06   }$  & 1.80e-03 & 1.30e-04 & -- & -- & --\\
  & 26.96   & -- & 2.54$^{+0.15   }_{-0.14   }$  & 1.95e-03 & 1.00e-04 & 1.02$^{+0.12   }_{-0.11   }$  & -- & --\\
  & 35.38   & -- & 2.82$^{+0.17   }_{-0.15   }$  & 2.28e-03 & 1.20e-04 & -- & -- & --\\
  & 46.75   & -- & 3.01$^{+0.28   }_{-0.25   }$  & 2.04e-03 & 1.10e-04 & 0.74$^{+0.12   }_{-0.10   }$  & -- & --\\
  & 59.39   & -- & 3.13$^{+0.29   }_{-0.27   }$  & 2.43e-03 & 1.30e-04 & -- & -- & --\\
  & 73.71   & -- & 4.01$^{+0.69   }_{-0.53   }$  & 1.99e-03 & 1.00e-04 & 0.47$^{+0.05   }_{-0.04   }$  & -- & --\\
  & 96.88   & -- & 3.33$^{+0.11   }_{-0.10   }$  & 9.19e-03 & 1.90e-04 & -- & -- & --\\
Hydra A & 19.55   & 4.84  & 2.92$^{+0.19   }_{-0.18   }$  & 3.17e-03 & 1.20e-04 & 0.48$^{+0.07   }_{-0.06   }$  & 1.16 & 999\\
  & 33.59   & -- & 3.02$^{+0.26   }_{-0.23   }$  & 3.87e-03 & 1.50e-04 & -- & -- & --\\
  & 49.47   & -- & 3.32$^{+0.28   }_{-0.26   }$  & 5.32e-03 & 1.50e-04 & 0.33$^{+0.04   }_{-0.03   }$  & -- & --\\
  & 72.69   & -- & 3.46$^{+0.33   }_{-0.28   }$  & 4.93e-03 & 1.50e-04 & -- & -- & --\\
  & 122.16  & -- & 3.71$^{+0.14   }_{-0.13   }$  & 1.26e-02 & 2.00e-04 & -- & -- & --\\
PKS0745-191 & 31.75   & 43.39 & 3.34$^{+0.19   }_{-0.18   }$  & 7.42e-03 & 2.80e-04 & 0.51$^{+0.07   }_{-0.08   }$  & 1.09 & 1472\\
  & 57.83   & -- & 5.31$^{+0.47   }_{-0.41   }$  & 9.03e-03 & 2.80e-04 & -- & -- & --\\
  & 91.85   & -- & 6.76$^{+0.68   }_{-0.60   }$  & 1.07e-02 & 3.00e-04 & 0.35$^{+0.05   }_{-0.04   }$  & -- & --\\
  & 147.42  & -- & 7.12$^{+0.56   }_{-0.51   }$  & 1.38e-02 & 3.00e-04 & -- & -- & --\\
  & 340.20  & -- & 8.57$^{+0.45   }_{-0.41   }$  & 2.52e-02 & 3.00e-04 & -- & -- & --\\
\enddata
\end{deluxetable}

\begin{deluxetable}{lrrrrrrrrrrr}
\tablewidth{0pt}
\tabletypesize{\scriptsize}
\tablecaption{Results for 2-T Deprojected, $\nh$ Fixed MEKAL Fits\label{tab:table2Tdeproj}}
\tablehead{
\colhead{Name} & \colhead{R$_{out}$ } & \colhead{$\nh$ } & \colhead{T$_{Hot}$ } & \colhead{Norm } & \colhead{$\sigma_{Norm}$} & \colhead{T$_{Cool}$ } & \colhead{Norm} & \colhead{$\sigma_{Norm}$} & \colhead{Z} & \colhead{$\chi^2$ } & \colhead{d.o.f}\\
\colhead{ } & \colhead{kpc} & \colhead{$10^{20}$ cm$^{-2}$} & \colhead{keV} & \colhead{Hot} & \colhead{Hot} & \colhead{keV} & \colhead{Cool} & \colhead{Cool} & \colhead{$Z_{\odot}$ } & \colhead{ } & \colhead{ }\\
\colhead{(1)} & \colhead{(2)} & \colhead{(3)} & \colhead{(4)} & \colhead{(5)} & \colhead{(6)} & \colhead{(7)} & \colhead{(8)} & \colhead{(9)} & \colhead{(10)} & \colhead{(11)} & \colhead{(12)} \\
}
\startdata
2A0335+096 & 12.44   & 18.11 & 2.04$^{+0.36   }_{-0.24   }$  & 1.48e-03 & 3.00e-04 & 1.06$^{+0.10   }_{-0.09   }$  & 6.02e-04 & 1.09e-03 & 0.90$^{+0.20   }_{-0.16   }$  & 1.32 & 1968\\
  & 19.07   & -- & 1.70$^{+0.09   }_{-0.08   }$  & 2.87e-03 & 3.40e-04 &   &   &  & -- & -- & --\\
  & 25.71   & -- & 1.97$^{+0.08   }_{-0.08   }$  & 6.18e-03 & 3.80e-04 &   &   &  & 0.81$^{+0.05   }_{-0.05   }$  & -- & --\\
  & 33.17   & -- & 2.21$^{+0.12   }_{-0.11   }$  & 5.27e-03 & 2.90e-04 &   &   &  & -- & -- & --\\
  & 42.29   & -- & 2.73$^{+0.15   }_{-0.14   }$  & 6.34e-03 & 2.70e-04 &   &   &  & 0.86$^{+0.04   }_{-0.07   }$  & -- & --\\
  & 54.31   & -- & 2.95$^{+0.17   }_{-0.16   }$  & 6.50e-03 & 2.50e-04 &   &   &  & -- & -- & --\\
  & 70.48   & -- & 3.62$^{+0.31   }_{-0.29   }$  & 5.76e-03 & 2.30e-04 &   &   &  & -- & -- & --\\
  & 90.80   & -- & 3.69$^{+0.23   }_{-0.22   }$  & 8.58e-03 & 2.30e-04 &   &   &  & 0.65$^{+0.05   }_{-0.05   }$  & -- & --\\
  & 165.84  & -- & 4.45$^{+0.11   }_{-0.11   }$  & 2.50e-02 & 4.00e-04 &   &   &  & -- & -- & --\\
A262 & 6.97    & 5.46  & 1.64$^{+0.22   }_{-0.17   }$  & 2.66e-04 & 9.70e-05 & 0.80$^{+0.03   }_{-0.03   }$  & 1.58e-04 & 1.80e-04 & 1.52$^{+1.34   }_{-0.57   }$  & 1.22 & 921\\
  & 13.74   & -- & 1.43$^{+0.04   }_{-0.04   }$  & 5.80e-04 & 7.30e-05 &   &   &  & 1.51$^{+0.13   }_{-0.18   }$  & -- & --\\
  & 21.12   & -- & 1.90$^{+0.05   }_{-0.11   }$  & 9.42e-04 & 7.10e-05 &   &   &  & -- & -- & --\\
  & 30.48   & -- & 2.01$^{+0.14   }_{-0.13   }$  & 1.06e-03 & 9.70e-05 &   &   &  & -- & -- & --\\
  & 40.64   & -- & 2.65$^{+0.29   }_{-0.26   }$  & 9.51e-04 & 1.74e-04 &   &   &  & 1.91$^{+0.91   }_{-0.27   }$  & -- & --\\
  & 59.76   & -- & 2.35$^{+0.07   }_{-0.05   }$  & 8.46e-03 & 2.50e-04 &   &   &  & 0.53$^{+0.06   }_{-0.05   }$  & -- & --\\
A2052 & 13.90   & 2.85  & 1.59$^{+0.54   }_{-0.27   }$  & 2.50e-04 & 7.80e-05 & 0.85$^{+0.12   }_{-0.07   }$  & 1.05e-04 & 1.90e-04 & 0.62$^{+0.14   }_{-0.05   }$  & 1.20 & 1543\\
  & 20.22   & -- & 1.40$^{+0.05   }_{-0.04   }$  & 1.56e-03 & 1.30e-04 &   &   &  & -- & -- & --\\
  & 26.96   & -- & 2.56$^{+0.13   }_{-0.14   }$  & 1.98e-03 & 8.00e-05 &   &   &  & 0.99$^{+0.06   }_{-0.11   }$  & -- & --\\
  & 35.38   & -- & 2.80$^{+0.17   }_{-0.15   }$  & 2.30e-03 & 5.00e-05 &   &   &  & -- & -- & --\\
  & 46.75   & -- & 3.01$^{+0.29   }_{-0.24   }$  & 2.03e-03 & 1.20e-04 &   &   &  & 0.75$^{+0.11   }_{-0.11   }$  & -- & --\\
  & 59.39   & -- & 3.13$^{+0.29   }_{-0.26   }$  & 2.43e-03 & 1.20e-04 &   &   &  & -- & -- & --\\
  & 73.71   & -- & 4.01$^{+0.67   }_{-0.53   }$  & 1.99e-03 & 1.00e-04 &   &   &  & 0.47$^{+0.05   }_{-0.04   }$  & -- & --\\
  & 96.88   & -- & 3.33$^{+0.11   }_{-0.10   }$  & 9.19e-03 & 1.90e-04 &   &   &  & -- & -- & --\\
\enddata
\end{deluxetable}


\begin{deluxetable}{lrlccccc}
\tablewidth{0pt}
\tabletypesize{\scriptsize}
\tablecaption{Fits to Temperature Profiles from Deprojected X-ray Spectra \label{tab:fit-temp}}
\tablehead{\colhead{Name} & \colhead{$N$}  & 
          \colhead{$T_{100}$} & \colhead{Slope} & \colhead{$\chi^2$} & \colhead{Prob} \\ 
\colhead{} & \colhead{}  & 
          \colhead{(keV)} & \colhead{} & \colhead{} & \colhead{} \\ }
\startdata
2A0335+096 & 9 & $4.0\pm0.2$ & $0.41\pm0.03$ & 3.5 & 0.75 \\
A133       & 4 & $4.1\pm0.3$ & $0.32\pm0.04$ & 1.0 & 0.31 \\
A262       &  6 & $3.1\pm0.2$ & $0.34\pm0.03$ & 2.0 & 0.57 \\
A496       & 6  & $5.1\pm0.7$ & $0.37\pm0.08$ & 0.8 & 0.86 \\
A1795       & 10 & $5.5\pm0.7$ & $0.30\pm0.09$ & 1.1 & 0.99 \\
A2029       & 13 & $7.2\pm0.6$ & $0.31\pm0.06$ & 3.3 & 0.97 \\
A2052       &  8 & $3.8\pm0.2$ & $0.47\pm0.05$ & 15.4 & 0.009 \\
Hydra A     &  5 & $3.7\pm0.3$ & $0.11\pm0.06$ & 0.2 & 0.92 \\
PKS0745-191 &  5 & $6.5\pm0.4$ & $0.34\pm0.05$ & 0.9 & 0.64 \\
\enddata
\end{deluxetable}

\begin{deluxetable}{lrccccccc}
\tablecaption{High Resolution Entropy Fit Results (Evenly Binned 5 Arcsecond Bins) \label{tab:fit-5arcsec}}
\tablehead{\colhead{Name} & \colhead{$N$} & \colhead{Radius} & \colhead{Method} & \colhead{$K_0$} & 
           \colhead{$K_{100}$} & \colhead{Slope} & \colhead{$\chi^2$} & \colhead{Prob} \\ 
           \colhead{} & \colhead{} & \colhead{(Mpc)} & \colhead{} & \colhead{(keV cm$^2$)} & 
           \colhead{(keV cm$^2$)} & \colhead{} & \colhead{} & \colhead{} \\ }
\startdata
2A0335& 38 &0.13&1&$5.438\pm0.34$&$135.4\pm2.5$&$1.41\pm0.026$& 61& 1e-3\\
&&&1&$=0$&$129.6\pm2.3$&$1.14\pm0.015$& 275& 0\\
&&&2&$7.240\pm0.33$&$134.7\pm2.5$&$1.49\pm0.028$& 74& 3e-05\\
&&&2&$=0$&$126.1\pm2.3$&$1.10\pm0.015$& 468& 0\\
A133& 25 &0.13&1&$11.82\pm 1.1$&$149.1\pm3.8$&$1.31\pm0.050$& 27& 0.09\\
&&&1&$=0$&$150.1\pm3.5$&$0.967\pm0.021$& 112& 2e-14\\
&&&2&$15.70\pm 1.0$&$145.1\pm3.8$&$1.39\pm0.054$& 33& 0.02\\
&&&2&$=0$&$144.8\pm3.5$&$0.906\pm0.020$& 185& 0\\
A1795& 31 &0.18&1&$14.57\pm 1.7$&$120.8\pm3.8$&$1.18\pm0.063$& 5.8& 0.99\\
&&&1&$=0$&$132.3\pm3.3$&$0.842\pm0.025$& 51& 0.003\\
&&&2&$19.97\pm 1.6$&$114.2\pm3.8$&$1.29\pm0.069$& 9.3& 0.99\\
&&&2&$=0$&$129.2\pm3.3$&$0.784\pm0.024$& 100& 2e-10\\
A2029& 28 &0.20&1&$6.347\pm 3.5$&$162.4\pm6.2$&$0.891\pm0.068$& 5.0& 0.99\\
&&&1&$=0$&$168.6\pm5.0$&$0.793\pm0.034$& 7.5& 0.99\\
&&&2&$10.18\pm 3.3$&$158.1\pm6.2$&$0.928\pm0.070$& 3.9& 0.99\\
&&&2&$=0$&$167.8\pm5.0$&$0.764\pm0.033$& 10& 0.99\\
A2052& 20 &0.070&1&$9.221\pm0.82$&$188.0\pm9.7$&$1.56\pm0.066$& 70& 2e-09\\
&&&1&$=0$&$153.5\pm5.9$&$1.10\pm0.029$& 176& 0\\
&&&2&$11.73\pm0.78$&$197.9\pm11.$&$1.70\pm0.075$& 87& 1e-14\\
&&&2&$=0$&$147.9\pm5.8$&$1.07\pm0.029$& 273& 0\\
A262& 31 &0.050&1&$1.022\pm0.38$&$230.0\pm8.8$&$1.11\pm0.028$& 154& 0\\
&&&1&$=0$&$218.2\pm6.6$&$1.05\pm0.013$& 162& 0\\
&&&2&$3.606\pm0.34$&$241.6\pm10.$&$1.20\pm0.031$& 190& 0\\
&&&2&$=0$&$195.9\pm5.9$&$0.979\pm0.013$& 290& 0\\
A496& 26 &0.080&1&$7.109\pm 1.0$&$175.0\pm9.6$&$1.20\pm0.065$& 5.8& 0.99\\
&&&1&$=0$&$150.8\pm6.6$&$0.902\pm0.029$& 39& 0.01\\
&&&2&$11.32\pm0.97$&$177.2\pm10.$&$1.31\pm0.073$& 8.2& 0.99\\
&&&2&$=0$&$133.0\pm5.8$&$0.788\pm0.027$& 90& 3e-10\\
Hydra A& 20 &0.10&1&$13.43\pm 1.1$&$98.74\pm3.9$&$1.22\pm0.078$& 2.2& 0.99\\
&&&1&$=0$&$92.18\pm2.8$&$0.684\pm0.024$& 72& 3e-09\\
&&&2&$14.06\pm 1.1$&$98.44\pm4.0$&$1.24\pm0.080$& 2.2& 0.99\\
&&&2&$=0$&$91.23\pm2.8$&$0.671\pm0.023$& 79& 2e-10\\
PKS0745& 32 &0.30&1&$5.908\pm 1.1$&$111.3\pm3.2$&$1.18\pm0.037$& 13& 0.98\\
&&&1&$=0$&$121.7\pm2.3$&$1.05\pm0.020$& 35& 0.16\\
&&&2&$12.37\pm 1.1$&$101.9\pm3.2$&$1.27\pm0.041$& 20& 0.78\\
&&&2&$=0$&$123.5\pm2.3$&$0.978\pm0.019$& 119& 3e-13\\

\enddata
\end{deluxetable}

\begin{deluxetable}{lrccccccc}
\tablecaption{High Resolution Entropy Fit Results (Evenly Binned 2.5" Annuli) \label{tab:fit-2.5arcsec}}
\tablehead{\colhead{Name} & \colhead{$N$} & \colhead{Radius} & \colhead{Method} & \colhead{$K_0$} & 
           \colhead{$K_{100}$} & \colhead{Slope} & \colhead{$\chi^2$} & \colhead{Prob} \\ 
\colhead{} & \colhead{} & \colhead{(Mpc)} & \colhead{} & \colhead{(keV cm$^2$)} & 
           \colhead{(keV cm$^2$)} & \colhead{} & \colhead{} & \colhead{} \\ }
\startdata
2A0335& 76 &0.13&1&$4.680\pm0.28$&$131.9\pm2.4$&$1.39\pm0.023$& 95& 0.02\\
&&&1&$=0$&$123.9\pm2.2$&$1.14\pm0.014$& 330& 0\\
&&&2&$6.562\pm0.27$&$131.7\pm2.5$&$1.48\pm0.025$& 115& 6e-3\\
&&&2&$=0$&$119.2\pm2.2$&$1.09\pm0.014$& 593& 0\\
A133& 49 &0.13&1&$11.19\pm0.99$&$149.7\pm3.7$&$1.32\pm0.046$& 37& 0.72\\
&&&1&$=0$&$146.8\pm3.3$&$0.974\pm0.019$& 129& 3.51186e-10\\
&&&2&$15.37\pm0.92$&$145.7\pm3.7$&$1.42\pm0.050$& 43& 0.47\\
&&&2&$=0$&$139.8\pm3.2$&$0.900\pm0.018$& 219& 0\\
A262& 61 &0.050&1&$1.174\pm0.30$&$240.0\pm9.6$&$1.15\pm0.027$& 141& 1e-09\\
&&&1&$=0$&$223.2\pm7.2$&$1.08\pm0.014$& 156& 3e-11\\
&&&2&$3.619\pm0.28$&$249.4\pm10.$&$1.24\pm0.030$& 163& 1e-14\\
&&&2&$=0$&$193.6\pm6.3$&$0.990\pm0.013$& 304& 0\\
A496& 51 &0.080&1&$6.771\pm 1.0$&$170.2\pm8.6$&$1.19\pm0.062$& 19& 0.99\\
&&&1&$=0$&$147.8\pm5.8$&$0.912\pm0.026$& 48& 0.40\\
&&&2&$11.44\pm0.93$&$172.4\pm9.6$&$1.32\pm0.070$& 23& 0.99\\
&&&2&$=0$&$129.9\pm5.1$&$0.793\pm0.023$& 105& 2e-06\\
A1795& 61 &0.18&1&$13.25\pm 1.7$&$121.5\pm3.1$&$1.17\pm0.053$& 15& 1.00\\
&&&1&$=0$&$132.6\pm2.6$&$0.876\pm0.020$& 56& 0.51\\
&&&2&$18.95\pm 1.5$&$114.7\pm3.1$&$1.28\pm0.058$& 20& 0.99\\
&&&2&$=0$&$130.1\pm2.6$&$0.824\pm0.019$& 111& 2e-05\\
A2029& 56 &0.20&1&$3.026\pm 2.1$&$164.9\pm4.3$&$0.863\pm0.044$& 13& 1.0\\
&&&1&$=0$&$167.7\pm3.7$&$0.814\pm0.025$& 15& 1.0\\
&&&2&$9.386\pm 1.9$&$157.9\pm4.2$&$0.930\pm0.047$& 10& 1.0\\
&&&2&$=0$&$166.2\pm3.7$&$0.766\pm0.024$& 27& 0.99\\
A2052& 41 &0.070&1&$7.270\pm0.81$&$179.3\pm7.3$&$1.48\pm0.054$& 116& 9e-11\\
&&&1&$=0$&$158.4\pm5.1$&$1.16\pm0.024$& 190& 0\\
&&&2&$9.648\pm0.77$&$186.2\pm8.2$&$1.60\pm0.060$& 142& 7e-15\\
&&&2&$=0$&$155.0\pm5.0$&$1.14\pm0.024$& 280& 0\\
Hydra A& 40 &0.10&1&$13.38\pm 1.0$&$98.58\pm3.4$&$1.23\pm0.068$& 6.7& 1.0\\
&&&1&$=0$&$92.28\pm2.4$&$0.705\pm0.020$& 91& 1e-06\\
&&&2&$14.08\pm 1.0$&$98.25\pm3.4$&$1.26\pm0.070$& 6.7& 1.0\\
&&&2&$=0$&$91.28\pm2.4$&$0.691\pm0.020$& 101& 4e-08\\
PKS0745& 65 &0.30&1&$4.831\pm0.78$&$109.4\pm2.3$&$1.17\pm0.027$& 33& 0.99\\
&&&1&$=0$&$117.4\pm1.8$&$1.06\pm0.017$& 69& 0.21\\
&&&2&$11.33\pm0.74$&$100.5\pm2.3$&$1.25\pm0.030$& 50& 0.79\\
&&&2&$=0$&$118.1\pm1.9$&$0.962\pm0.015$& 241& 0\\

\enddata
\end{deluxetable}

\begin{deluxetable}{lrccccccc}
\tablecaption{Entropy Fit Results (Evenly Binned 2.5" Annuli, Temperature from best-fit deprojected temp) \label{tab:fit-2.5arcsec-tdeproj}}
\tablehead{\colhead{Name} & \colhead{$N$} & \colhead{Radius} & \colhead{Method} & \colhead{$K_0$} & 
           \colhead{$K_{100}$} & \colhead{Slope} & \colhead{$\chi^2$} & \colhead{Prob} \\ 
\colhead{} & \colhead{} & \colhead{(Mpc)} & \colhead{} & \colhead{(keV cm$^2$)} & 
           \colhead{(keV cm$^2$)} & \colhead{} & \colhead{} & \colhead{} \\ }
\startdata
2A0335& 76 &0.13&3&$2.673\pm0.19$&$123.7\pm2.2$&$1.27\pm0.019$& 114& 7e-3\\
&&&3&$=0$&$115.7\pm2.0$&$1.09\pm0.013$& 270& 0\\
A133& 49 &0.13&3&$6.949\pm0.79$&$152.8\pm3.1$&$1.15\pm0.031$& 44& 0.41\\
&&&3&$=0$&$150.0\pm2.9$&$0.955\pm0.013$& 105& 8e-07\\
A262& 61 &0.050&3&$1.214\pm0.27$&$206.4\pm8.2$&$1.08\pm0.024$& 82& 0.01\\
&&&3&$=0$&$190.1\pm6.4$&$1.00\pm0.013$& 101& 3e-3\\
A496& 51 &0.080&3&$3.755\pm0.97$&$159.1\pm7.9$&$1.05\pm0.053$& 17& 0.99\\
&&&3&$=0$&$145.3\pm5.9$&$0.891\pm0.023$& 28& 0.98\\
A1795& 61 &0.18&3&$11.75\pm 1.7$&$116.1\pm4.2$&$1.19\pm0.069$& 7.9& 1.0\\
&&&3&$=0$&$127.2\pm3.6$&$0.872\pm0.023$& 36& 0.98\\
A2029& 56 &0.20&3&$4.783\pm 1.2$&$144.7\pm5.5$&$1.03\pm0.059$& 5.5& 1.0\\
&&&3&$=0$&$148.9\pm5.3$&$0.867\pm0.029$& 16& 0.99\\
A2052& 46 &0.080&3&$3.351\pm0.53$&$141.8\pm6.0$&$1.33\pm0.043$& 58& 0.03\\
&&&3&$=0$&$130.7\pm5.1$&$1.15\pm0.027$& 93& 9e-06\\
Hydra A& 40 &0.10&3&$10.90\pm 1.0$&$99.18\pm2.8$&$1.13\pm0.058$& 6.4& 1.0\\
&&&3&$=0$&$96.67\pm2.2$&$0.736\pm0.018$& 71& 4e-3\\
PKS0745& 65 &0.30&3&$5.053\pm0.57$&$110.5\pm1.9$&$1.18\pm0.022$& 38& 0.98\\
&&&3&$=0$&$119.6\pm1.6$&$1.05\pm0.013$& 106& 3e-3\\

\enddata
\end{deluxetable}

\end{document}